\documentclass[12pt]{article}
\usepackage{amsmath}
\usepackage{graphicx}
\usepackage{enumerate}
\usepackage{natbib}
\usepackage{url} 
\usepackage[colorlinks=true,allcolors=blue]{hyperref}
\usepackage{url}
\usepackage{graphicx}
\usepackage{amsmath, amssymb}
\usepackage{bbold}
\usepackage{mathbbol}
\usepackage{algorithm}
\usepackage{algorithmic}
\usepackage{enumitem}
\usepackage{float}
\usepackage{xcolor}
\usepackage{bm} 
\usepackage{setspace}
\usepackage{multirow}
\usepackage{lscape}
\usepackage{rotating}
\usepackage{adjustbox}
\usepackage{booktabs}
\usepackage{ragged2e}
\usepackage{caption}
\usepackage{subcaption}
\usepackage{float}
\def\*#1{\mathbf{#1}}
\def\+#1{\amsmathbb{#1}}
\def\^#1{\mathbb{#1}}
\DeclareSymbolFontAlphabet{\amsmathbb}{AMSb}%

\newcommand{\blind}{1}

\usepackage{bnumexpr}

\usepackage{algorithm}

\usepackage{hyperref}

\usepackage{amsthm}

\newtheorem{proposition}{Proposition}

\begin{document}

\def\spacingset#1{\renewcommand{\baselinestretch}%
{#1}\small\normalsize} \spacingset{1}


\if1\blind
{
  \title{\bf Multivariate Scalar on Multidimensional Distribution Regression}
  \author{Rahul Ghosal$^1$, Marcos Matabuena$^{2,3}$\\
     $^{1}$ Department of Epidemiology and Biostatistics, University of South Carolina \\
$^{2}$Department of Biostatistics, Harvard University, Boston, USA\\
$^{3}$ University of Santiago de Compostela, A Coruña, Spain\\
}
  \maketitle
} \fi

\if0\blind
{
  \bigskip
  \bigskip
  \bigskip
  \begin{center}
    {\LARGE\bf Functional proportinal hazards mixture cure model}
\end{center}
  \medskip
} \fi

\bigskip
\begin{abstract}
We develop a new method for multivariate scalar on multidimensional distribution regression. Traditional approaches typically analyze isolated univariate scalar outcomes or consider unidimensional distributional representations as predictors. However, these approaches are sub-optimal because: i) they fail to utilize the dependence between the distributional predictors: ii)  neglect the correlation structure of the response. To overcome these limitations, we propose a multivariate distributional analysis framework that harnesses the power of multivariate density functions and multitask learning. We develop a computationally efficient semiparametric estimation method for modelling the effect of the latent joint density on multivariate response of interest. Additionally, we introduce a new conformal algorithm for quantifying the uncertainty of regression models with multivariate responses and distributional predictors, providing valuable insights into the conditional distribution of the response. We have validated the effectiveness of our proposed method through comprehensive numerical simulations, clearly demonstrating its superior performance compared to traditional methods. The application of the proposed method is demonstrated on tri-axial accelerometer data from the National Health and Nutrition Examination Survey (NHANES) 2011-2014 for modelling the association between cognitive scores across various domains and distributional representation of physical activity among older adult population. Our results highlight the advantages of the proposed approach, emphasizing the significance of incorporating complete spatial information derived from the accelerometer device.

\end{abstract}

\noindent%
{\it Keywords:}  Distributional Data Analysis; Multivariate Analysis; Scalar on Distribution Regression; Physical activity; NHANES; Cognitive Score.
\vfill

\newpage
\spacingset{1.9} 
\section{Introduction}
\label{sec:intro4}

\noindent Advancements in technology have opened up new opportunities to capture clinical information at an unprecedented level of resolution using modern wearables and smartphones. These devices allow for the collection of vast quantities of data streams, encompassing various physiological signals such as energy expenditure (a proxy for physical activity), heart rate, continuously monitored blood glucose, and more. Exploring and comprehensively understanding these data streams and their underlying distributional patterns hold tremendous potential for gaining deeper insights into human behaviors and their impact on human health as well as disease progression. By analyzing the distributional characteristics of these data streams, researchers can uncover valuable information that can inform personalized interventions and targeted disease management strategies.
The collection of data using wearable devices often involves monitoring patients in free-living conditions, which presents significant methodological challenges. Traditional methods like time series analysis or functional data analysis \citep{Ramsay05functionaldata} are not directly applicable in these cases due to the lack of standardized conditions and varying durations of stochastic processes across different patients.

\noindent In the analysis of wearable data, two common approaches have been followed. The first approach involves analyzing specific moments of the entire time series by calculating, for example, the sample mean or another moments \citep{gait2020vr}. While this approach simplifies interpretation, it has drawbacks such as the loss of information when only considering the first moments of the time series and omit the local dynamic of time series. Another approach is to use compositional vector metrics, which quantitatively measure the proportion of time spent in specific value ranges \citep{janssen2020systematic}. For instance, in the case of diabetes, the proportion of time spent in the hypoglycemic range ($<$70 mg/dL) can be measured \citep{sherr2013reduced}. However, this approach also has limitations. Discretizing the data into intervals can lead to a loss of information and introduce subjectivity in the statistical analysis. Determining the optimal intervals for analysis becomes challenging, and the results may depend on the chosen thresholds.

\noindent These limitations highlight the need for alternative methods that can overcome the loss of information and subjectivity introduced by discretization or threshold-based approaches. It is crucial to develop robust statistical techniques that can handle the complexities of free-living data and effectively exploit the functional information collected by the devices at different time scales. Distributional data analysis (DDA) which uses functional distributional representations \citep{gait2020rv,matabuena2020glucodensities} provide a powerful framework that surpasses the limitations of traditional data analysis methods. They enable us to capture and utilize a more comprehensive set of information regarding various clinical outcomes of patients. By incorporating the entire distribution of the data, these representations offer a functional extension of compositional metrics, allowing for more nuanced interpretation and analysis.
DDA has already gained significant traction and showcased its versatility through a myriad of applications across numerous scientific domains, encompassing but not limited to digital health \citep{matabuena2020glucodensities,gait2020rv,ghosal2022scalar,matabuena2023distributional}, neuroimaging \citep{zhu2021distributional,tang2023differences} and various other fields such as sleep research. In their comprehensive review, \cite{petersen2021modeling} delved into the latest advancements in DDA, with a particular emphasis on leveraging densities as a primary analytical tool.

\begin{figure}[H]
\begin{center}
\begin{tabular}{ll}
\includegraphics[width=.5\linewidth , height=.55\linewidth]{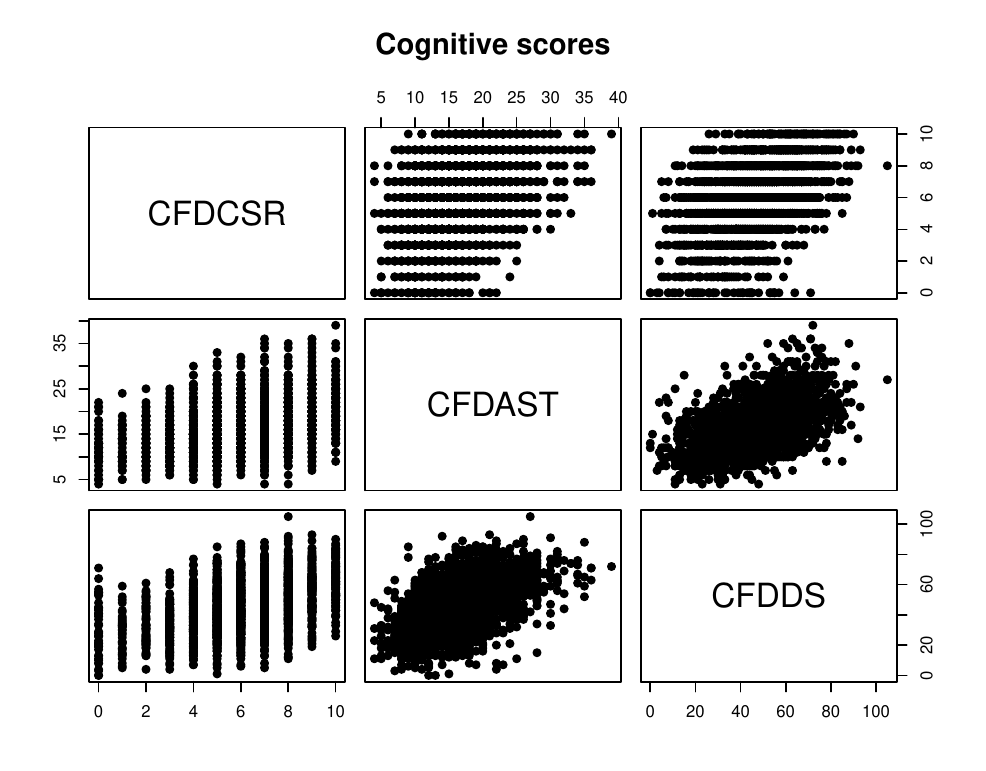} &
\includegraphics[width=.5\linewidth , height=.5\linewidth]{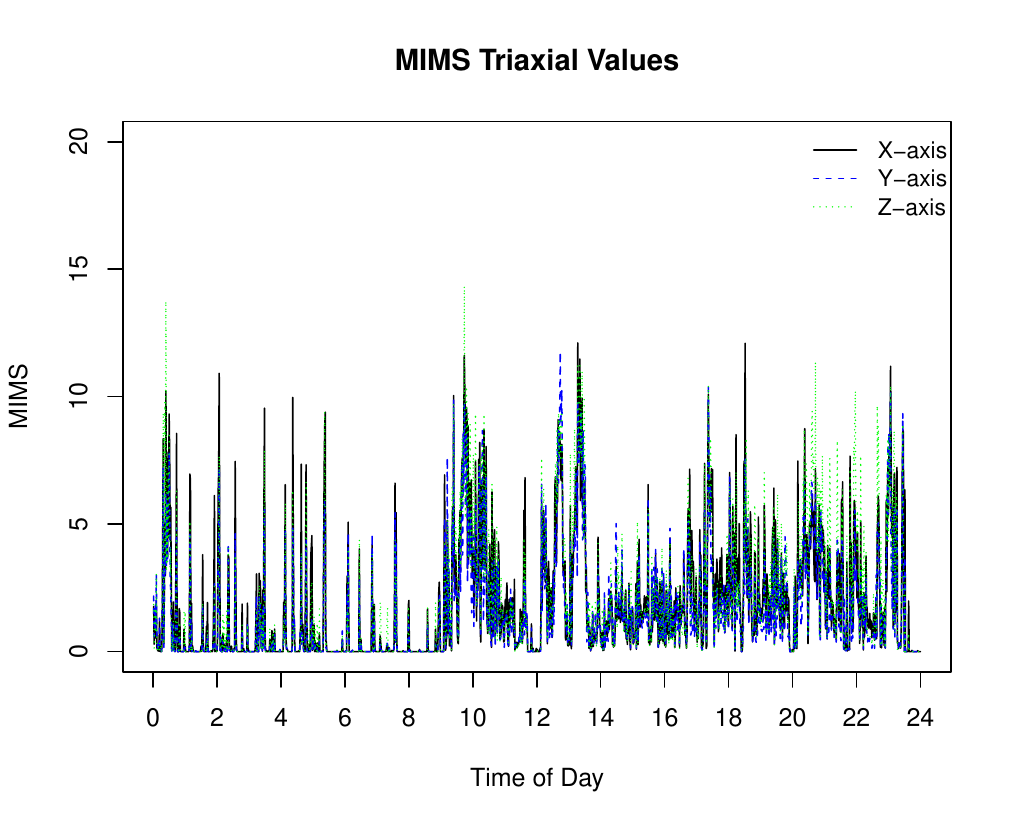}
\end{tabular}
\end{center}
\caption{Left: Scatterplot of cognitive scores Right : Observed Triaxial MIMS profile of a representative NHANES participant on a sample day.}
\label{fig:fig1m}
\end{figure}

In this paper, as a motivating application we consider modelling the association between cognitive scores from three different domains (CFDCSR: delayed recall, CFDAST: animal fluency test, CFDDS: digit symbol substitution test (DSST)) in the National Health and Nutrition Examination Survey (NHANES) 2011- 2014 and objectively measured physical activity data collected by tri-axial accelerometer among the older adult population.  Figure \ref{fig:fig1m} displays the observed correlation matrix between the cognitive scores along with the tri-axial minute-level MIMS values for a representative participant on a sample day. Previous research have explored the dose-response relationship between the cognitive scores and summary level PA metrics such as average or peak monitor-independent movement summary (MIMS) units \citep{zheng2023dose} or association with the average diurnal pattern of PA \citep{antonsdottir202324}. However, using the complete distributional information in such continuous concurrent streams of data could potentially provide a better understanding into the association between cognition and PA.

\noindent In the field of statistical modeling and machine learning, regression analysis plays a crucial role in understanding the relationships between variables and making predictions. 
Distributional regression models offer the flexibility to accommodate distributional characteristics in either the predictors, the response variable, or both. Within the realm of modeling scalar responses of interest, such as health outcomes, researchers have put forth numerous proposals for distributional regression models. Functional compositional methods \citep{petersen2016functional,hron2016simplicial} map densities to a proper Hilbert space $\mathcal{L}^2$ and then use existing functional regression approaches for modelling scalar outcomes. Other research in this area have used different distributional representations e.g., transformed densities, densities or quantile functions and used them in a scalar-on-distribution regression framework \citep{talska2021compositional,gait2020rv,matabuena2021distributional,ghosal2022scalar,meunier2022distribution,matabuena2023distributional,tang2023direct}. Nevertheless, these current scalar-on-distribution regression methods have primarily examined univariate scalar outcomes in isolation or concentrated on uni-dimensional distributional predictors that can be represented through univariate entities like transformed densities or quantile functions. However, when confronted with correlated outcomes and multiple distributional predictors, relying solely on marginal or univariate approaches can result in a loss of statistical efficiency. These approaches fail to adequately consider the correlation among the outcomes and the interdependence among the distributional predictors. To address these limitations and improve the accuracy of the analysis, it becomes crucial to develop comprehensive methodologies.

In this article, we develop a multivariate scalar on multi-dimensional distributional regression (MSOMDR) that aims to model multivariate scalar responses, e.g., cognitive scores, based on joint-distribution of multiple and simultaneously observed continuous streams of data. The proposed method combines the tools of multivariate regression and distributional regression to provide a more complete understanding of the relationships between distributional predictors and multivariate outcomes. The primary advantage of MSOMDR is its ability to incorporate the joint dependence between multiple distributional predictors in modeling the multivariate response. Moreover, by simultaneously considering the joint effects of the distributional predictors on the multivariate response, MSOMDR provides a more comprehensive way to account for the correlation between the outcomes.  

\subsection{Paper contributions: }

Our article makes several significant contributions to the fields of distributional data analysis and wearable data analysis. The key contributions are summarized below:

\begin{enumerate}
    \item We propose a multivariate scalar-on-multidimensional distributional regression model that extends existing multitask learning approaches to distributional regression models. This model directly captures the effect of subject-specific joint density on the multivariate outcome of interest. The new model overcomes one of the most critical challenges in using density as a predictor in regression modeling, namely the direct interpretation of the statistical association with the response at each point of the density domain in terms of a $\beta(\cdot)$ functional coefficient. The new interpretability method measures statistical association in the raw monitor intensities, overcoming the limitation of using quantiles, where the same quantile is individualized across subjects and it is not easy to compare the covariate effect. This representation marks an important advance in the field of compositional data analysis from an interpretability perspective.
    
    \item We introduce a computationally efficient estimation approach using splines to handle multivariate density responses as predictors in regression modeling. This approach employs a simple modeling trick that allows us to accommodate density functions as predictors that are statistical objects taking values in a non-linear space \cite{petersen2019frechet}.
    \item We present a new conformal inference method specifically designed for constructing prediction regions of the multivariate response with a multidimensional distributional predictor. This method provides non-asymptotic guarantees and offers a distribution-free framework for uncertainty quantification in MSOMDR.
\end{enumerate}

Finally, in the NHANES study case:

\begin{enumerate}
    \item We demonstrate the practical utility of the proposed MSOMDR method through an application to modeling the cognitive score of patients incorporating physical activity information measured by tridimensional accelerometer data. Specifically, we show that the new tridimensional distributional approach can improve the predictive capacity in this modeling task compared to traditional summary univariate metrics that include univariate distributional representations.
    
    \item Compared to existing approaches, the results highlight the importance of incorporating spatial information in data analysis and the significance of modeling the conditional distribution to incorporate uncertainty quantification techniques, such as conformal prediction, in this scientific problem.
\end{enumerate}
 
\subsection{Paper outline:}
The subsequent sections of this article are organized as follows. In Section 2, we introduce the modeling framework, delve into the concept of multivariate distributional representation and present the multivariate scalar on multidimensional distributional regression model. Moving on to Section 3, we outline the estimation approach for the MSOMDR model and introduce a conformal inference algorithm that facilitates the construction of prediction regions. We present the results from numerical simulations in Section 4 to evaluate the performance of the proposed method and provide insightful comparisons with existing scalar-on-distribution regression approaches.  In Section 5, we demonstrate the practical application of our proposed method using the NHANES data. Finally, in Section 6, we conclude our article with a concise discussion of our contributions and outline potential avenues for extending this work.

\section{Methodology}
\subsection{Modelling Framework and Multidimensional distributional representation}
We consider the scenario, where there are repeated subject-specific measurements of a multivariate 
distributional predictor $\*Z=(Z_1,Z_2,\ldots,Z_d) \in \mathbb{R^d}$ along with a multivariate scalar outcome of interest $\*Y=(Y_1,Y_2,\ldots,Y_K)\in \mathbb{R^k}$ 
and several scalar covariates $X_j$, $j=1,2,\ldots,q$, which serves as relevant confounders. Let us denote the observed data for subject $i$ as $D_i=\{\*Y_i,X_{i1},\ldots X_{iq}, \*Z_{il};l=1,\ldots,m_{i}\}$, for subject $i=1,\ldots,n$. Here $m_{i}$ denotes the number of repeated observations of the distributional predictor $\*Z$ for subject $i$. We assume that $\*Z_{il}$ $(l=1,\ldots,n_{1i}) \sim P_{\*Z_i}$ with $ F_{i\*Z}(\*z)$ being the corresponding subject-specific multivariate cumulative distribution function (c.d.f), where $F_{i\*Z}(\*z)=P(Z_{i1}\leq z_1,\ldots,Z_{id}\leq z_d)$. In practice, we don't observe the latent subject-specific distribution, rather have observations $\*Z_{il}$ $(l=1,\ldots,m_{i})$ from that underlying distribution. We further assume that $\*Z_{il}$ are absolutely continuous, having a density function $f_{i\*Z}(\*z)$ corresponding to the c.d.f $F_{i\*Z}(\*z)$.  It is possible to use kernel density estimation or density based approaches \citep{petersen2021modeling} to estimate the latent subject specific joint distribution $P_{\*Z_i}$ and map them to a element of a reproducing kernel Hilbert space (RKHS). In the next section, we illustrate how our model formulation and estimation approach can bypass the estimation of the latent density and directly models the multivariate scalar outcomes of interest based on the realizations from the underlying distribution.  

\subsection{Multivariate Scalar on Multidimensional Distributional Regression Model}
We propose the following distributional regression model, modelling the multivariate scalar outcome $Y_{ik}$ ($k=1,\ldots,K$) based on observations $\*Z_{il}$ from the multidimensional distribution 
$P_{\*Z_i}$ and scalar covariates $X_{ij}$, $j=1,2,\ldots,q$. We refer to this model as a multivariate scalar on multidimensional distribution regression (MSOMDR).
\begin{eqnarray}
Y_{ik}=\*X_{i}^T\bm\gamma_k+\int_{\*s\in \mathbb{R^d}} \beta_{k}(\*s)P_{\*Z_i}(d\*s)+\epsilon_{ik}. \label{mod 1}
   \end{eqnarray}
We assume $\epsilon_{ik}$ are i.i.d. mean zero random error 
which are uncorrelated with $X_{ij}$ and $\*Z_{il}$s.  
The coefficient $\bm\gamma_k$ captures the scalar effect of the confounders $\*X_{i}^T=(X_{i1},\ldots, X_{iq})$ on the outcome $Y_{ik}$. The distributional effect of $\*Z_{il}$ (and latent $P_{\*Z_i}$) on $Y_{ik}$ is captured by $\beta_{k}(\*s)$. Previously \cite{tang2023direct} has considered univariate and unidimensional distributional analogue of the above model from a Bayesian perspective, the proposed MSOMDR serves as a multivariate and multidimensional generalization. Although the above model is developed for a general dimension $d$ of the distributional predictor, the main focus of this paper will be scenarios where $d=2$ or $d=3$ for notational simplicity, matching with the motivating real data application considered in this paper. 


\section{Estimation}
The MSOMDR model for a 3-dimensional distributional predictor is given by
\begin{eqnarray}
Y_{ik}=\*X_{i}^T\bm\gamma_k+\int_{\mathcal{W}}\int_{\mathcal{V}}\int_{\mathcal{U}}\beta_{k}(u,v,w) P_{\*Z_i}(du dv dw)+\epsilon_{ik}.
\end{eqnarray}
Since $\*Z_{il}$ are absolutely continuous, having a density function $f_{i\*Z}(\cdot)$, the above model can be reformulated as
\begin{eqnarray}
Y_{ik}=\*X_{i}^T\bm\gamma_k+\int_{\mathcal{W}}\int_{\mathcal{V}}\int_{\mathcal{U}}f_{i\*Z}(u,v,w)\beta_{k}(u,v,w)du dv dw+\epsilon_{ik}.
\end{eqnarray}
The above model resembles a multivariate scalar-on-functional regression model \citep{reiss2017methods}, where the functional predictor is multi-dimensional \citep{marx2005multidimensional}. Note that, the latent density $f_{i\*Z}(\cdot)$ is not directly observed, rather we observe $\*Z_{il}$ coming from this latent density (distribution). Instead of using a density estimation approach which could be problematic, especially in higher dimensions, we directly use the observations $\*Z_{il}$ as in \cite{tang2023direct} to approximate $E_{P_{\*Z_i}}(\beta_k(\*z))=\int_{\mathcal{W}}\int_{\mathcal{V}}\int_{\mathcal{U}}f_{i\*Z}(u,v,w)\beta_{k}(u,v,w)du dv dw \approx \frac{1}{m_{i}}\sum_{l=1}^{m_i}\beta_{k}(\*Z_{il})$. Based on this approximation, we can further reduce model (3) to
\begin{eqnarray}
Y_{ik}=\*X_{i}^T\bm\gamma_k+\frac{1}{m_{i}}\sum_{l=1}^{m_i}\beta_{k}(\*Z_{il})+\epsilon_{ik}, \hspace{3 mm} k=1,\ldots,K
\end{eqnarray}
which bypasses the need for estimating the latent subject-specific densities (or distributional representations).
The unknown parameters of interest are the scalar coefficients $\bm\gamma_k$ and the multi-dimensional distributional effect $\beta_{k}(\cdot)$, which need to be estimated. 

The multi-dimensional link function $\beta_k(u,v,w)$ is modelled using tensor decomposition in terms of tensor product of univariate cubic B-spline basis functions as   $$\beta_k(u,v,w)=\sum_{f=1}^{N_U}\sum_{g=1}^{N_V} \sum_{h=1}^{N_W}\theta_{f,g,h}^{k}B_{fU}(u) B_{gV}(v) B_{hW}(w)=\*W(u,v,w)^T\bm\theta_K,$$ where $\{B_{fU}(u)\}_{f=1}^{N_U}$, $\{B_{gV}(v)\}_{g=1}^{N_V}$ and $\{B_{hW}(w)\}_{h=1}^{N_W}$  are sets of known B-spline basis function over $u,v,w$ respectively (taken to be same across all coefficient functions). We denote the stacked vector \\$\{B_{fU}(u)B_{gV}(v)B_{hW}(w)\}_{f=1,g=1,h=1}^{N_U,N_V,N_W}$ of length $N_0=N_UN_VN_W$ as $\*W(u,v,w)^T$ and similarly $\bm\theta_k$ is the stacked coefficient vector $\{\theta_{f,g,h}^k\}_{f=1,g=1,h=1}^{N_U,N_V,N_W}$.
Plugging in this expression in model (4) we have,
\begin{eqnarray}
Y_{ik}=\*X_{i}^T\bm\gamma_k+\frac{1}{m_{i}}\sum_{l=1}^{m_i}\*W(\*Z_{il})^T\bm\theta_k+\epsilon_{ik}\notag \\
=\*X_{i}^T\bm\gamma_k+\*W_i^T\bm\theta_k+\epsilon_{ik},\hspace{3 mm} k=1,\ldots,K,
\end{eqnarray}
where $\*W_i^T=\frac{1}{m_{i}}\sum_{l=1}^{m_i}\*W(\*Z_{il})^T$ and $\*Z_{il}=(Z_{il1},Z_{il2},Z_{il3})$. Now stacking the response across all the outcomes we can write,
\begin{eqnarray}
\*Y_i&=&\+U_i\bm\gamma+\+V_i\bm\theta+\bm\epsilon_{i}, \hspace{3 mm} i=1,\ldots,n, \label{MSODR1}
\end{eqnarray}
where $\+U_i=I_{K\times K}\otimes\*X_{i}^T$, $ \+V_i=I_{K\times K}\otimes\*W_{i}^T$, $\*Y_i=(Y_{i1},Y_{i2},\ldots,Y_{iK})$, $\bm\epsilon_{i}=(\epsilon_{i1},\epsilon_{i2},\ldots,\epsilon_{iK})$, $\bm\theta=(\bm\theta_1^T,\bm\theta_2^T,\ldots,\bm\theta_K^T)^T$ and $\bm\gamma=(\bm\gamma_1^T,\bm\gamma_2^T,\ldots,\bm\gamma_K^T)^T$. We alternatively denote $\bm\theta_k$ to be $\bm\theta_k^T=(\theta_{1k},\theta_{2k},\ldots,\theta_{N_0k})$ which will be used in latter sections. Hence the MSOMDR model reduces to a multivariate linear model on the tensor product of the basis matrices. Since, the dimension of the basis coefficients $\bm\theta$ gets exponentially large with increasing dimension $d$, we use a multi-task learning approach using group penalized regression \citep{breheny2015group}
for estimating the model parameters $\bm\theta$ and $\bm\gamma$ which simultaneously introduces shrinkage and correlation among the multivariate outcomes due to their dependency on similar predictors \citep{pecanka2019modeling}. In particular, we minimize the following penalized least square criterion,
\begin{equation}\hat{\bm\psi}=(\hat{\bm\gamma},\hat{\bm\theta})=\underset{\bm\gamma,\bm\theta}{\text{argmin}}\hspace{1 mm}\sum_{i=1}^{n}||\*Y_i-\+U_i\bm\gamma-\+V_i\bm\theta||_{2}^{2}+n\sum_{l=1}^{N_0}P_{MCP,\lambda,\phi}(||\bm\theta_{\ell.}||_2),
\label{est:FOSR}
\end{equation}
where $\bm\theta_{\ell.}^T=(\theta_{\ell1},\theta_{\ell2},\ldots,\theta_{\ell K})$, which captures the effect of $W_{l}$ on the multivariate outcome $\*Y$. The group minimax concave penalty (MCP) \citep{zhang2010nearly} $P_{MCP,\lambda,\phi}(||\bm\theta_{\ell.}||_2)$ on the basis coefficients is defined in the following way,
\begin{equation*}
P_{MCP,\lambda,\phi}(||\bm\theta_{\ell.}||_2)=
\begin{cases}
\lambda||\bm\theta_{\ell.}||_2-\frac{||\bm\theta_{\ell.}||_2^2}{2\phi}\hspace{1.4 cm} \text{if $||\bm\theta_{\ell.}||_2\leq \lambda\phi$}.\\
.5\lambda^2\phi \hspace{3.19 cm} \text{if $||\bm\theta_{\ell.}||_2>\lambda\phi$}.\\
\end{cases}
\end{equation*}
The number of basis functions $N_U,N_V,N_W$ work as a tuning parameter of the above minimization problem, controlling the smoothness of the coefficient functions $\beta_k(u,v,w)$ and they are implicitly controlled using a truncated basis approach \citep{Ramsay05functionaldata,fan2015functional}. The tuning parameters and the penalty parameter $\lambda$ are chosen in a data-driven way using V-fold ($V=5$ used in this article) cross-validation approach on the cross-validated sum of squared errors corresponding to model (6). The tuning parameter $\phi$ is set $3$ for the MCP, based on the recommendations by  \cite{zhang2010nearly}. We have used the {\tt grpreg} package \citep{bre2015} in R \citep{Rsoft} for implementation of the above optimization (7) using a group descent algorithm. Once the parameters $\hat{\bm\theta}$ and $\hat{\bm\gamma}$ are estimated, the estimates of the distributional coefficient functions are given by $\hat{\beta}_k(u,v,w)=\sum_{f=1}^{N_U}\sum_{g=1}^{N_V} \sum_{h=1}^{N_W}\hat{\theta}_{f,g,h}^{k}B_{fU}(u) B_{gV}(v) B_{hW}(w), \hspace{2mm} k=1,2,\ldots K$.

\subsection{Conformal Prediction with the MSOMDR Model}
Conformal inference, a versatile framework for uncertainty quantification  \citep{hammouri2023uncertainty}  in both supervised and unsupervised settings, has emerged as a pivotal domain in contemporary statistical research \citep{vovk2005algorithmic,lei2018distribution,romano2019conformalized,angelopoulos2023conformal}. For practical purposes, conformal prediction techniques is a general methodology to provide predictions regions and characterize the conditional distribution between  two random variables, $Y$ and $X$. Conformal prediction offer the following key advantages:
\begin{itemize}
    \item  They provide prediction regions that are agnostic to the specific predictive regression model employed.

    \item  They offer non-asymptotic guarantees concerning marginal coverages under general exchangeability assumptions.

    \item  They deliver predictive regions that are fundamentally non-parametric in nature.

\end{itemize}

In this paper, we introduce a novel conformal inference algorithm \citep{diquigiovanni2022conformal} designed to accommodate multivariate responses and multidimensional distributional representations. Our approach leverages the geometric properties of the supremum norm as a reference for these extensions, simplify  the final computation of prediction regions for multivariate responses. An exceptional feature of employing the supremum norm in constructing these prediction regions is its ability to enhance clinical interpretability. By representing these regions as hypercubes, healthcare professionals can gain valuable insights, facilitating the establishment of clinical diagnostic thresholds and defining the limits of predictive models \citep{matabuena2022kernel}. Similar to our approach, in \cite{young2020nonparametric}, the authors use a supremum norm geometric approach but with depth bands to define tolerance regions of type $I$ and without considering covariates. In our conformal inference framework, we specifically focus on the conditional case with covariates to establish tolerance regions of type $II$ (tolerance reference regions in expectation). See \cite{li2008multivariate} for a discussion about this topic.

From a practical perspective of explanation of our proposed algorithm, we consider a multivariate response, denoted as $\mathbf{Y} \in \mathcal{Y} = \mathbb{R}^{K}$, and assume that the vector of random errors $\bm{\epsilon} = (\epsilon_1, \ldots, \epsilon_K)$ in the MSOMDR model (1) satisfies $\mathbb{E}(\bm{\epsilon} | \mathbf{X}, P_{\mathbf{Z}}) = 0$. In essence, we link the regression model, represented by the function $m(\cdot)$, to the conditional mean estimator. Our primary objective is to construct a prediction region, denoted as $\mathcal{C}^{\alpha}(\mathbf{X}, P_{\mathbf{Z}}) \subset \mathbb{R}^{K}$, which provides a confidence level of $\alpha$ for capturing the response variable based on the conditional mean regression function. Specifically, our aim is to achieve $P(\mathbf{Y} \in \mathcal{C}^{\alpha}(\mathbf{X}, P_{\mathbf{Z}} )) = 1 - \alpha$. To identify such a region, additional criteria are required, including the minimization of volume within or constraining the regions to a particular geometry, as in our case with the supremum norm. We refer to this population-level prediction region as the `oracle prediction region'. In practice, we observe a random sample $\mathcal{D}_{n} = {(\mathbf{X}_{i}, P_{\mathbf{Z}_i}, \mathbf{Y}_{i})}_{i=1}^{n}$, where each subject's data is independently and identically distributed or at least exchangeable. Employing conformal inference techniques, we can construct prediction regions with non-asymptotic guarantees of the type  as $P(\mathbf{Y} \in \widehat{\mathcal{C}}_{n}^{\alpha}(\mathbf{X}, P_{\mathbf{Z})}) \geq 1-\alpha$. As the sample size $n$ approaches infinity, convergence towards the oracle prediction region is achieved. 


To develop the  new conformal inference algorithm that will be computationally efficient, we introduce a data-splitting strategy  that in literature of conformal inference is known as split-conformal  
\citep{papadopoulos2002inductive}. In our application and distributional models, using this computational strategy is necessary, since the large number of participants that we analyze elevate the cost to fit the distributional models involving multidimensional-splines. Specifically, we  partition the sample set $\mathcal{D}_{n}$ into three distinct and independent random samples: $\mathcal{D}_{train1}$, $\mathcal{D}_{train2}$, and $\mathcal{D}_{calibration}$. The sample $\mathcal{D}_{train1}$ serves as the primary training set, which we employ to train our model and learn the underlying relationship between the predictor variables and the response variable in terms of estimating the regression function $m(\*X,P_{\*Z}))=(m_1(\*X,P_{\*Z}),\ldots,m_K(\*X,P_{\*Z}))^T$, where $m_k(\*X,P_{\*Z}))=\*X^T\bm\gamma_k+\int_{\*s\in \mathbb{R^d}} \beta_{k}(\*s)P_{\*Z}(d\*s)$. The dataset $\mathcal{D}_{\text{train2}}$ serves as a secondary training set, which is utilized to estimate the modulation vector for calculating the nonconformity scores \citep{diquigiovanni2022conformal}. In this study, we have employed the standard deviation as the modulation function to capture the local variability of the data, denoted as $sd(\cdot)$. To estimate the standard deviation regression function, we employed the residuals with respect to the estimator of the regression function $m(\cdot)$, evaluated in $\mathcal{D}_{\text{train2}}$, using the conditional mean estimator once again.

 \begin{algorithm} [ht]
		\caption{Conformal prediction algorithm for MSOMDR}
\begin{enumerate}
			\item Estimate the function $m(\cdot,\cdot)$, by $\hat{m}(\cdot)$ based on the random sample $\mathcal{D}_{train1}$ using the proposed MSOMDR estimation method.
   \item For all $i\in \mathcal{D}_{train2}$, evaluate $\hat{m}(X_i,P_{Z_i})$, define ${r}_{ik}= |Y_{ik}- \hat{m}_{k}(X_i,P_{Z_i})|$, and with the random sample $\{((X_i,P_{Z_i}),\*r_{i})\}_{i\in \mathcal{D}_{train2}}$, obtain $\hat{s}_{k}(X_i,P_{Z_i},\*r_i)=  \hat{sd}(r_{ik})$, denote $\*s^T=(\hat{s}_{1},\ldots,\hat{s}_{K})$.
			
\item For all $i \in \mathcal{D}_{calibration}$, define $R_{i}= \sup_{k=1,\dots,K}\frac{ |Y_{ik}-\hat{m}_{ik}(X_i,P_{\*Z_i})|}{\hat{s}_{k}}$. 
 
			\item Estimate the empirical distribution $\widetilde{G}^{*}(t)
			=  \frac{1}{|\mathcal{D}_{calibration}|}  \sum_{i\in \mathcal{D}_{calibration}} 1\{R_i\leq t\}$ and denote by $ \hat{{q}}_{1-\alpha}$ the empirical quantile of level $1-\alpha$.
			\item Return $\widehat{\mathcal{C}} _{n}^{\alpha}(\*X,P_{\*Z})= [\hat{m}(\*X,P_{\*Z})-  \hat{q}_{1-\alpha}\hat{\*s}, \hat{m}(\*X,P_{\*Z})+   \hat{{q}}_{1-\alpha}\hat{
   \*s}]$

  \end{enumerate}

  \label{alg:metd1}
	\end{algorithm}

The sample $\mathcal{D}_{calibration}$ plays a crucial role in the conformal inference process. We utilize this set to calibrate the algorithm and determine the appropriate confidence levels or significance thresholds required for constructing the prediction region based on the nonconformity scores. Calibration ensures that the resulting regions adequately capture the desired uncertainty level and maintain proper coverage probabilities in the non-asymptotic regime. 

Algorithm \ref{alg:metd1} presents the core steps of the proposed conformal prediction algorithm. 

\begin{proposition}

\noindent For any function estimator of the regression function $m(\cdot,\cdot)$, $\hat{m}(\cdot,\cdot)$, the prediction region $\widehat{\mathcal{C}}_{n}^{\alpha}(\*X,\*P_{Z})$ for a new observation $(\*X,P_{\*Z})$ defined by the Algorithm  \ref{alg:metd1} satisfy:
 \begin{equation*}
P(\*Y\in \widehat{\mathcal{C}}_{n}^{\alpha}(\*X,P_{\*Z}))\geq 1-\alpha
\end{equation*}
\begin{proof}
See Appendix A in Supplementary Material.
\end{proof}
\end{proposition}

\section{Simulation Studies}
\label{simul}
In this Section, we investigate the performance of the proposed estimation and conformal prediction method for MSOMDR via simulations. To this end, we consider the following data generating scenarios.

\subsection{Data Generating Scenarios}
\subsection*{Scenario A1: MSOMDR, Estimation}
We consider the MSOMDR model given by,
\begin{eqnarray}
Y_{ik}=\*X_{i}^T\bm\gamma_k+\int_{\mathcal{V}}\int_{\mathcal{U}}\beta_{k}(u,v) P_{\*Z_i}(du dv)+\epsilon_{ik}, \hspace{2mm} k=1,2. \label{sim1}
\end{eqnarray}

The scalar predictors $\*X_{i}^T \in \mathbb{R}^2$  are generated independently from a bivariate normal distribution with mean $\bm\mu=(0,0)$ and covariance matrix $\^\Sigma=\begin{pmatrix}1 & 0.5\\ 0.5 & 1\end{pmatrix}$ and the corresponding scalar coefficients are $\bm\gamma_1=(1,3)$ and $\bm\gamma_2=(2,4)$. We observe realizations $\*Z_{il}$s $\in \^R^2$ ($l=1,\ldots m_i$) from the subject-specific multidimensional distribution $P_{\*Z_i}$. In particular, $\*Z_{il}\sim N(\bm\mu_i,C_i*\^\Sigma_0)$, where $\^\Sigma_0=\begin{pmatrix}1 & 0.3\\ 0.3 & 1\end{pmatrix}$, $\bm\mu_i=(\mu_{1i},\mu_{2i})$, $C_i\sim Unif (1,3)$ and both $\mu_{1i},\mu_{2i}$ follows a $N(0,1)$ distribution. The distributional effects are taken to be $\beta_{1}(u,v)=\frac{1}{2}(u^2+v^2)$ and $\beta_{2}(u,v)=\frac{1}{3}(u+4v+2uv)$ respectively. The residuals $\epsilon_{ik}$ are independently sampled from $N(0,1)$ for each $k$. Note that the dependence between the multivariate outcome is introduced through their dependence on the common set of predictors. We assume that $m_{i}=m=1000$ observations $\*Z_{il}$ are available for each subject. Sample size $n=500,1000,2000$ is considered for this data generating scenario, out of which $80\%$ is used for model training and estimation and the rest $20\%$ is used as a test set for evaluating out-of-sample prediction performance. We use 100 Monte-Carlo replications from the above scenario for model assessment.

\subsection*{Scenario A2: MSOMDR, Conformal Prediction}
We generate data from the same MSOMDR model (\ref{sim1}) as in scenario A1 above. This scenario will be used to assess the performance of the proposed conformal prediction algorithm. For the sampling design we consider the following scheme. Three sets of total  Sample size $n=500,1000,2000$ is considered, out of which $80\%$ is used for model training and calibration and the rest $20\%$ is used for evaluating coverage of the the prediction region. The samples for training and calibration are randomly partitioned into $\mathcal{D}_{train1}$, $\mathcal{D}_{train2}$, and $\mathcal{D}_{calibration}$ with equal probability. We again assume that $m_{i}=m=1000$ observations $\*Z_{il}$ are available for each subject. 100 Monte-Carlo replications from the above scenario are used for model assessment.

\subsection{Simulation Results}
\subsection*{Performance under scenario A1:}
\begin{figure}[H]
\centering
\includegraphics[width=0.7\linewidth , height=0.6\linewidth]{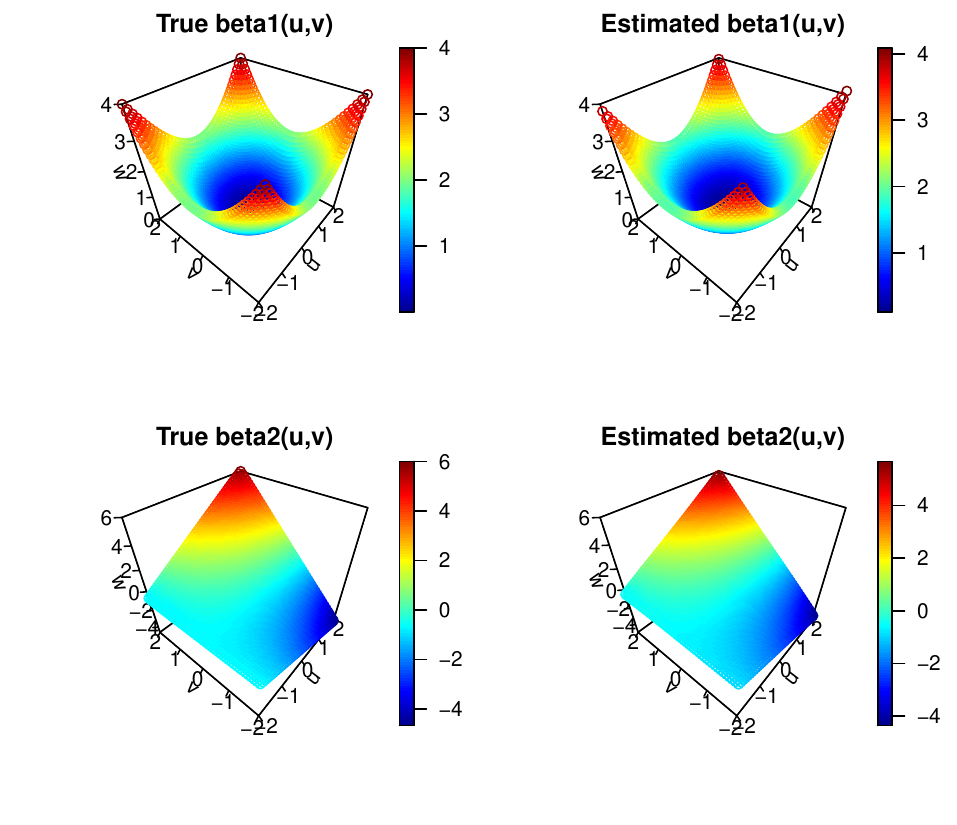}
\caption{Displayed are the true (left) and Monte Carlo mean of the estimated distributional effects (right) $\beta_1(u,v)$,$\beta_2(u,v)$, scenario A1, n=1000.}
\label{fig:fig2}
\end{figure}

We evaluate the performance of our proposed method in terms of estimation accuracy and out-of-sample prediction accuracy. The Monte Carlo (MC) mean estimates
of $\beta_1(u,v)$ and $\beta_2(u,v)$ (averaged over the 100 MC
replications) are displayed in Figure \ref{fig:fig2} for $n = 1000$ and over a grid of $u$ and $v$ within the support of the distributional predictors. 

We observe the true coefficient surfaces are closely captured by their corresponding estimates indicating a satisfactory performance of the proposed method in estimating the unknown distributional effects. The performances of $n=500,2000$ are illustrated in the Supplementary Figure S1 and S2. Supplementary Figure S3 displays the distribution of the $L^2$ loss $L_j^{b}=\{\int_{\mathcal{V}} \int_{\mathcal{U}} \{\hat{\beta}_j^{b}(u,v)-\beta_j(u,v)\}^2dudv\}^{\frac{1}{2}}$ ($b=1,2,.\ldots,100$) between the true and estimated distributional coefficients across the MC replications. The estimation accuracy can be noticed to gradually improve with increasing sample size as expected.

\begin{figure}[H]
\centering
\includegraphics[width=0.8\linewidth , height=0.6\linewidth]{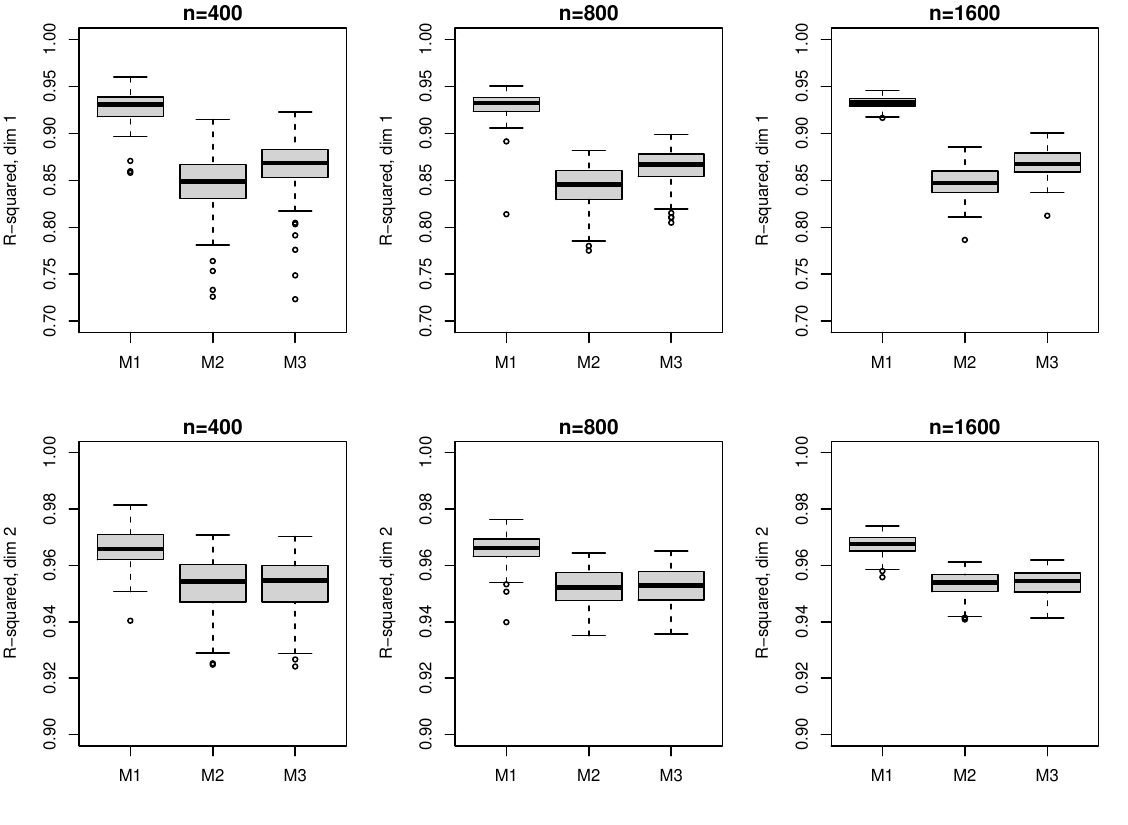}
\caption{Displayed are the R-squared value ($R^2_k,k=1,2$ in top and bottom row) in test data, for the three competing methods M1, M2, M3 across three sample sizes, scenario A1.}
\label{fig:fig3}
\end{figure}

The out-of-sample prediction performance of the proposed method is compared with two competing modelling approaches i) using uni-dimensional summary metrics (mean) of $\*Z_{il}$, $\mu_{\*Z_i}=(\mu_i^1,\mu_i^2)$ in a multivariate linear regression model (denoted as $M2$) and ii) using an additive scalar-on-quantile function regression \citep{gait2020rv} framework (SOQFR)
based on $\*Z_{il}$ in both the dimensions (denoted as $M3$). This approach use $Q_{i1}(p)$ (obtained from $Z_{il}^1$s ) and $Q_{i2}(p)$ (obtained from $Z_{il}^2$s) as distributional predictors in a multivariate SOQFR model. We use R-squared in test data for each $Y_k$ ($k=1,2$) as a measure of out of sample prediction performance, defined as $R^2_k=1-\frac{\sum_{i\in test} (Y_{ik}-\hat{Y}_{ik})^2}{\sum_{i\in test} (Y_{ik}-\Bar{Y}_{k})^2}, k=1,2$. Figure \ref{fig:fig3} displays the distribution of $R^2_k,k=1,2$ for MSOMDR (denoted as M1) and the other two competing methods across the three training sample sizes $n=400,800,1600$. It can be observed that the proposed MSOMDR method yields a higher test R-squared values for both the outcomes, illustrating the superiority of the proposed method. The gain is particularly substantial for outcome 1, where the true distributional effects are highly nonlinear.

\subsection*{Performance under scenario A2:}

\begin{figure}[H]
\centering
\includegraphics[width=0.4\linewidth , height=0.4\linewidth]{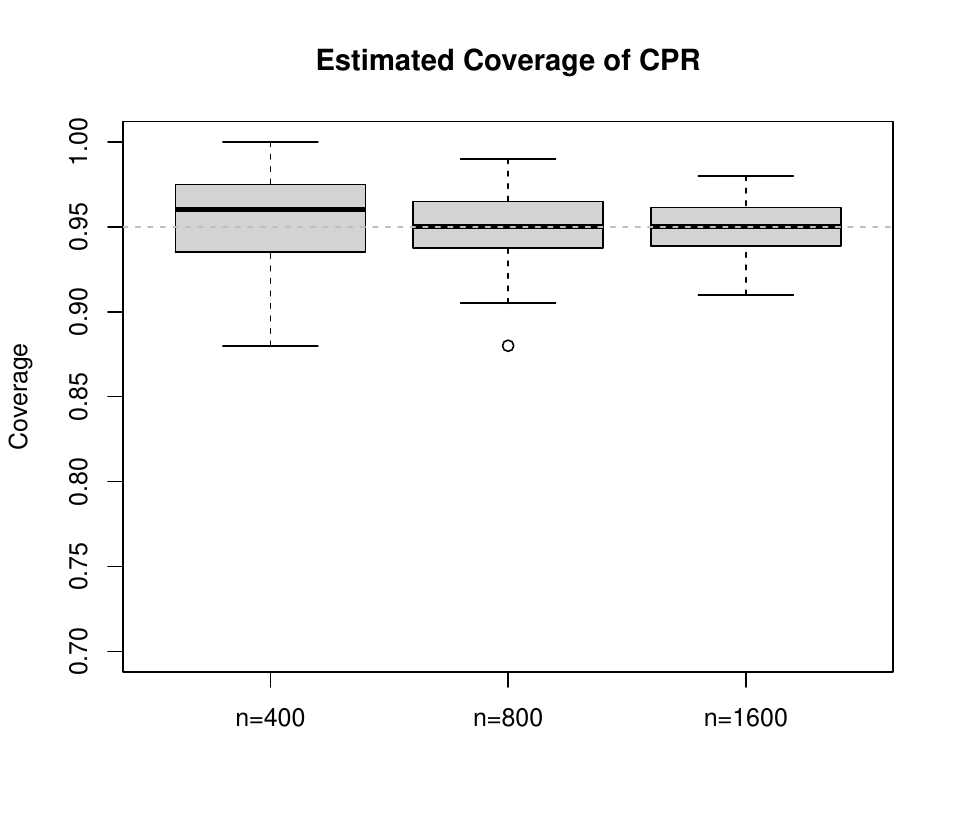}
\caption{Displayed are estimated coverage of the conformal prediction region (CPR) for the three different sample sizes, scenario A2.}
\label{fig:fig5}
\end{figure}

Next, we evaluate the performance of the proposed conformal prediction algorithm in terms of estimated coverage. The confidence level $\alpha$ is set to $0.05$. Algorithm 1, presented in this paper is used to obtain the multivariate prediction region, for each subject in test data and averaged across all the test-subjects to evaluate coverage. The distribution of the estimated coverage of the multivariate prediction region across the Monte Carlo replications are shown in Figure \ref{fig:fig5}, for the three training sample sizes. It can be observed that the estimated coverage is close to the nominal coverage value of $0.95$ across the sample size, illustrating  a satisfactory performance of the proposed algorithm. A reduction in the variability of the estimated coverage can also be noticed for higher sample sizes.

\section{Data Application: NHANES 2011-2014}
In this section, we apply the proposed MSOMDR method to tri-
axial accelerometer data from the National Health and Nutrition Examination Survey (NHANES) 2011-2014. The NHANES provide a broad range of descriptive health and nutrition statistics and is a nationally representative sample of the non-institutionalized US population.  In NHANES 2011-2014, 
acceleromtry data was collected using the wrist-worn ActiGraph GT3X+ accelerometer (ActiGraph of Pensacola, FL). Participants were asked to wear the physical activity monitor continually for seven full days (midnight to midnight) and remove it on the morning of the 9th day. We focus on the minute level and 2011-2014 accelerometer data, released in 2021, which reports individuals’ acceleration in Monitor Independent Movement Summary (MIMS) unit, an open-source, device-independent universal summary metric  \citep{john2019open}.
MIMS is available at each minute as a triaxial summary (MIMS triaxial value for the minute: sum of X,Y,Z axis MIMS) and also individually for X,Y, and Z axis. As mentioned in the introduction, the objective of our analysis is to quantify the association between cognitive scores across three different domains and multidimensional distributional representation of physical activity (from three different axes) among the older adult population in USA.

In particular, we consider cognitive scores from three different domains, i) CFDCSR: (delayed recall), ii) CFDAST: (animal fluency test), iii) CFDDS: (digit symbol substitution test (DSST)). Cognitive scores are extremely useful to examine the association of cognitive functioning with the medical conditions and other risk factors and tracking cognitive decline in aging population \citep{anderson2007cognitive}. NHANES 2011-2014 provide the above mentioned cognitive scores though a series of assessments on participants aged 60 years and older. A total of 1947 adults aged 60–80 years with available cognitive scores, physical activity data (physical
activity monitoring available at least ten hours per day for
at least four days) and covariate information (age, Gender) were included in our analysis. Supplementary Table S1 presents the descriptive statistics of the sample.

\begin{figure}[H]
\begin{center}
    \begin{tabular}{ll}
        \scalebox{0.5}{\includegraphics{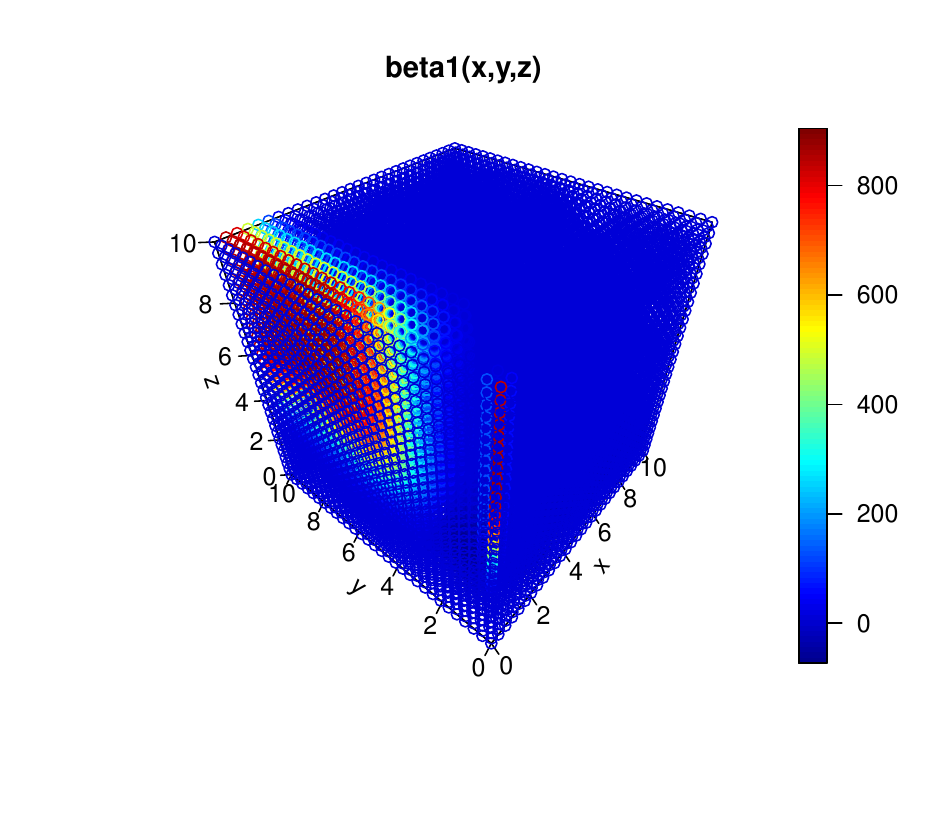}} &
 \scalebox{0.5}{\includegraphics{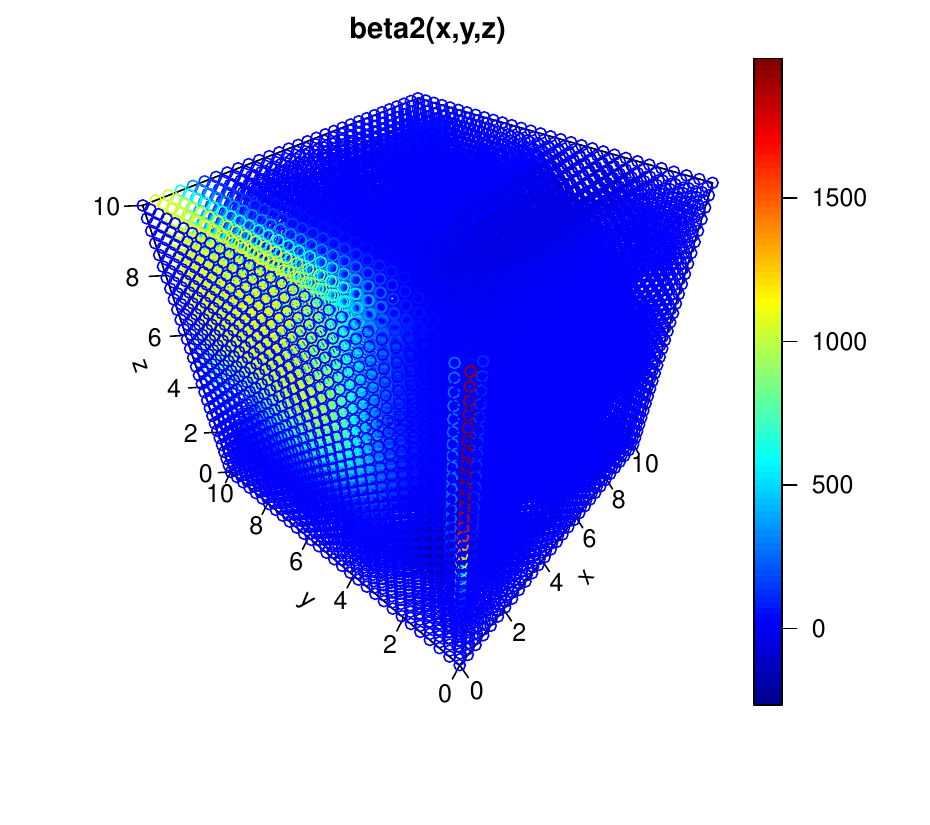}}\\
         \scalebox{0.5}{\includegraphics{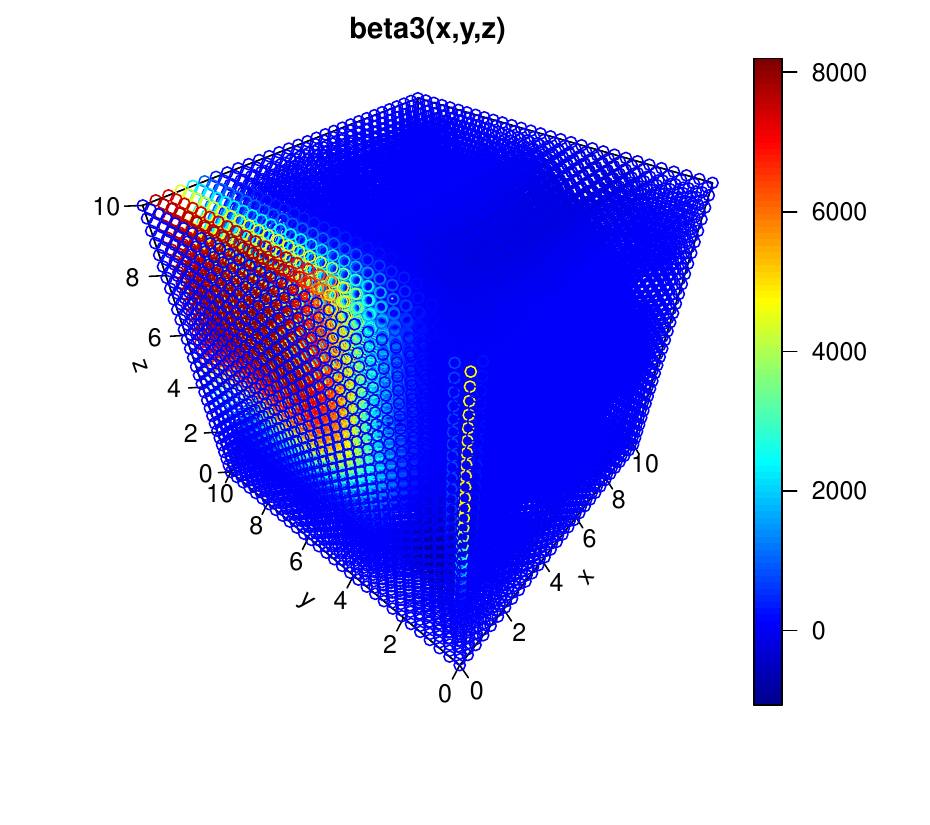}} &
 \\
    \end{tabular}
\end{center}
\caption{Estimated 3-dimensional distributional effects $\beta_1(x,y,z)$, $\beta_2(x,y,z)$ and $\beta_3(x,y,z)$ of joint density of 3-dimensional MIMS values on CFDCSR (top left), CFDAST (top right) and CFDDS (bottom left) cognitive scores respectively.}
\label{fig:fig4rd}
\end{figure}

We consider the following MSOMDR framework proposed in this paper to model multivariate cognitive scores based on multi-dimensional distributional representation of physical activity (MIMS).

\begin{eqnarray}
Y_{ik}=age_i\gamma_{1k}+G_i \gamma_{2k}+\int_{\mathcal{Z}}\int_{\mathcal{Y}}\int_{\mathcal{X}}\beta_{k}(x,y,z) P_{\*M_i}(dx dy dz)+\epsilon_{ik}, \hspace{2mm} k=1,2,3.
\end{eqnarray}
Here $Y_{ik}, k=1,2,3$ represents the cognitive scores  CFDCSR, CFDAST and CFDDS respectively for subject $i$. $G_i$ ($G_i=1$ for female, $G_i=0$ for male) is indicator variable for person's Gender. The subject specific tri-dimensional distribution of MIMS is represented by $P_{\*M_i}$. The distributional effects $\beta_{k}(x,y,z), k=1,2,3$ capture the three dimensional effect of latent subject-specific PA density at x,y,z values of MIMS (PA) on the three cognitive scores. We use cubic B-spline basis with varying number of basis functions in $x,y,z$ direction to model $\beta_{k}(x,y,z)$. The knots are placed at quantiles of data to yield better data coverage. The optimal number of basis functions based on 5-fold cross validation is chosen to be $N_X=9,N_X=12,N_X=12$ respectively. 
The estimated effects of age and Gender (female) on
the there cognitive scores are given by $\hat{\bm\gamma}_{1}^T=(-0.08,-0.12,-0.52)$ and $\hat{\bm\gamma}_{2}^T=(0.86,-0.14,6.06)$ respectively. Figure \ref{fig:fig4rd} displays the estimated distributional effects  $\hat{\beta}_{k}(x,y,z), k=1,2,3$ by the MSOMDR model.

Interestingly, we observe that for all three cognitive scores, a higher frequency (density) in higher $Y$ and $Z$- axis MIMS is associated with a higher CFDCSR (delayed recall),
CFDAST (animal fluency test) and CFDDS (digit symbol substitution TEST) score, indicating a better cognition in all the three domains. These cognitive scores have been used in large-scale screenings and epidemiologic studies \citep{grundman2004mild,proust2007sensitivity} and have been shown to discriminate between mild cognitive impairment and Alzheimer's disease \citep{henry2004verbal}. Since the Actigraph accelerometer is wrist-placed, higher intensity movements along Y and Z axis is indicative of higher intensity movement PA behaviour like: walking, jogging, upstairs, downstairs, standing of the participants \citep{javed2020analyzing}. Our results are among the first to confirm the dose-response relationship between physical activity and cognitive function among older adults \citep{zheng2023dose} in a nationally representative US population and particularly highlights the importance of high intensity activities which result in higher Y,Z-axis MIMS accumulation for a better cognitive functioning.

The proposed MSODR framework is more general and encompasses effects of any distributional features which are computed from X, Y and Z axis MIMS values, like vector magnitude (VM=$\sqrt{X^2+Y^2+Z^2}$) or total activity count (TAC, a summary from composite triaxial MIMS), or triaxial MIMS ($X+Y+Z$). The proposed MSOMDR method is compared with the following modelling approaches for comparison of out-of-sample prediction performance: i) using uni-dimensional summary metrics TAC (total activity count per day, based on triaxial composite MIMS) in a multivariate linear regression model (denoted as $CM_1$) and ii) using an additive scalar-on-quantile function regression \citep{gait2020rv} framework (SOQFR)
based on X,Y,Z axis MIMS (denoted as $CM_2$). This approach use $Q_{iX}(p)$ (obtained from $X_{il}$s, subject-specific X axis MIMS), $Q_{iY}(p)$ (obtained from $Y_{il}$s, subject-specific Y axis MIMS) and $Q_{iZ}(p)$ (obtained from $Z_{il}$s, subject-specific Z axis MIMS) as distributional predictors in a multivariate SOQFR model. iii) using subject-specific distribution of composite triaxial MIMS, $Q_{iA}(p)$ in a multivariate SOQFR model (denoted as $CM_3$).  

\begin{figure}[H]
\centering
\includegraphics[width=0.9\linewidth , height=0.4\linewidth]{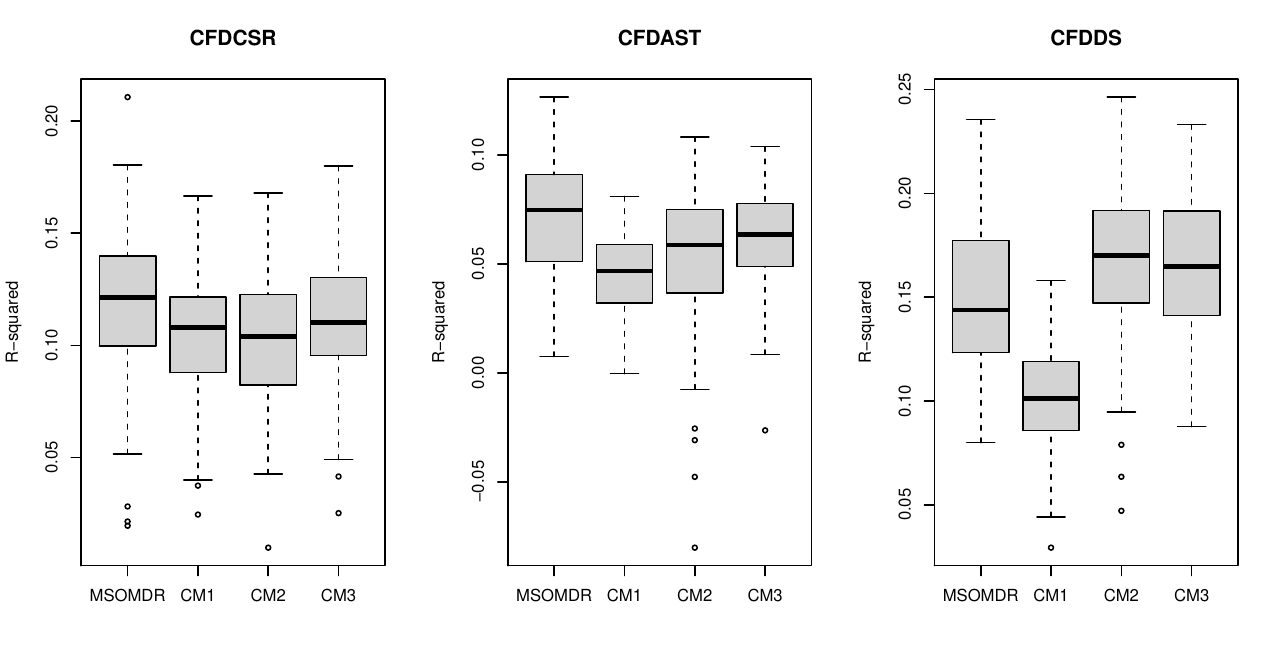}
\caption{Displayed are the R-squared value ($R^2_k,k=1,2,3$ for thre three cognitive scores in the test data from the MSOMDR and three competing methods CM1, CM2, CM3.}
\label{fig:figrsqdat}
\end{figure}

We use $80\%$ of the samples for training and $20\%$ for testing and the whole experiment is repeated $B=100$ times. Figure \ref{fig:figrsqdat} displays the distribution of the out-of-sample R-squared values in the test data for the proposed MSOMDR and the three competing methods. We observe that the proposed MSOMDR method explains a higher percentage of variation in the CFDCSR and the CFDAST cognitive scores compared to the TAC and additive univariate distributional approaches. For the CFDDS cognitive score, the performance of the additive SOQFR approach or the triaxial MIMS based approach is marginally better, illustrating that in this case an additive or triaxial MIMS based modelling might be sufficient.

\begin{figure}[H]
\begin{center}
\begin{tabular}{ll}
\includegraphics[width=.55\linewidth , height=.55\linewidth]{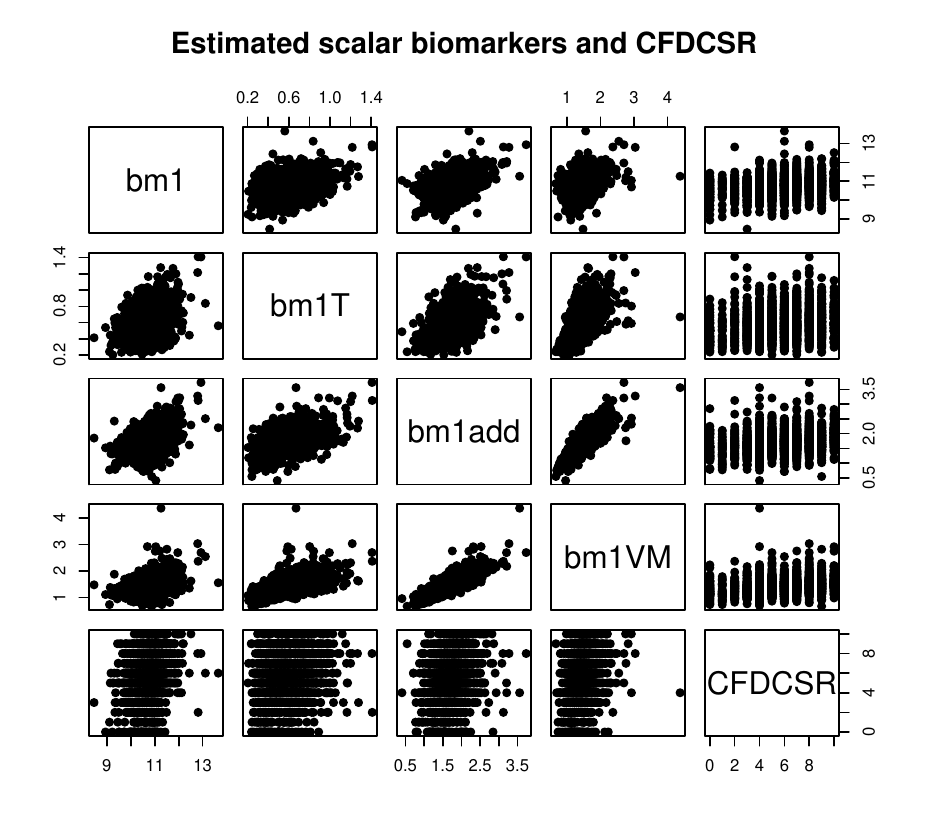} &
\includegraphics[width=.55\linewidth , height=.55\linewidth]{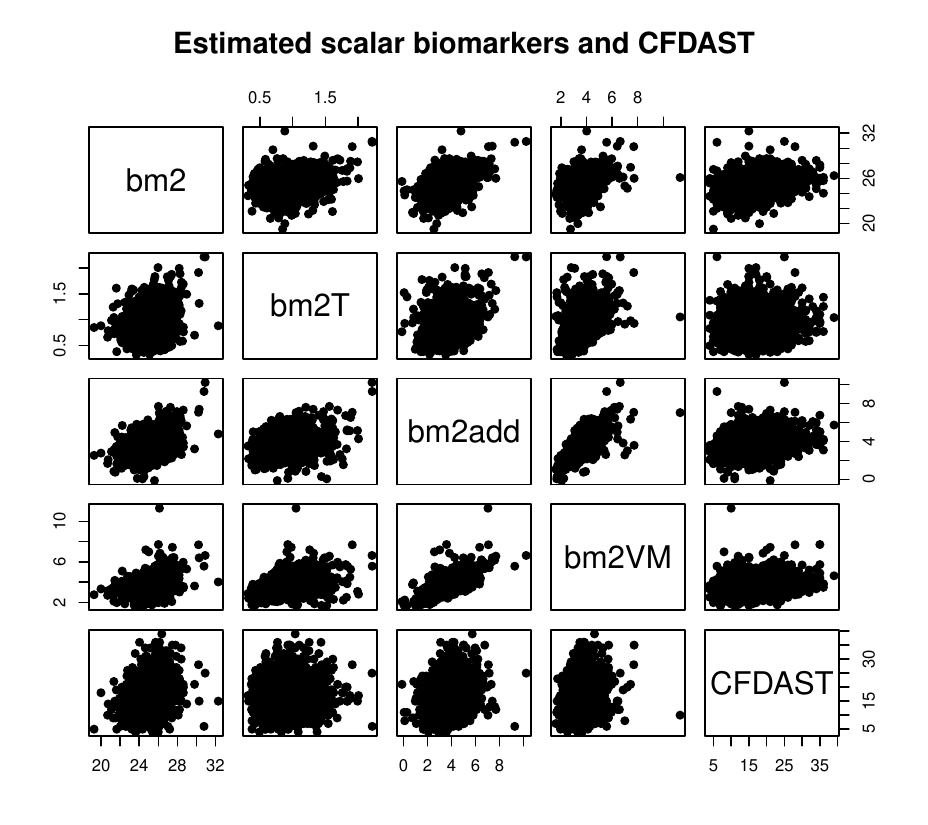}\\
\includegraphics[width=.55\linewidth , height=.55\linewidth]{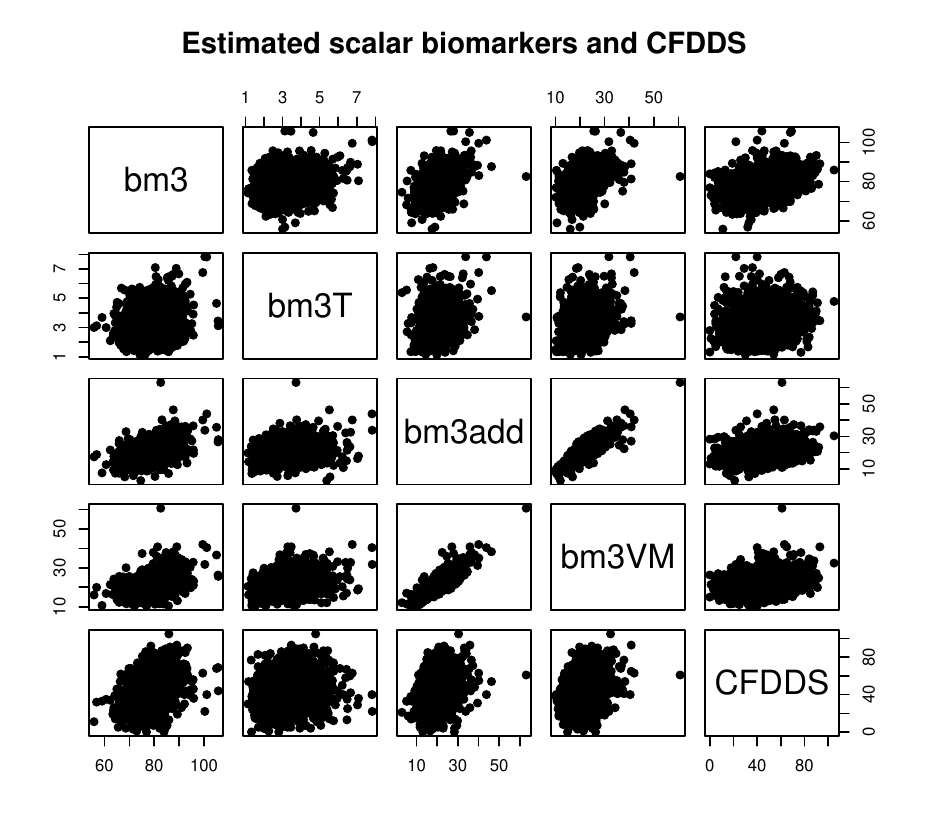} 
\end{tabular}
\end{center}
\caption{Scatterplots for estimated distributional biomarkers from the MSOMDR and the three competing approaches for the three cognitive scores considered in the NHANES application.}
\label{fig:figbm}
\end{figure}

We can use the estimated distributional coefficients from the MSOMDR model to create interpretable scalar biomarkers for the three cognitive scores. For example, based on 
$\hat{\beta}_{k}(x,y,z)$, we define the following biomarkers $bm_{ki} = \int_{\mathcal{Z}}\int_{\mathcal{Y}}\int_{\mathcal{X}}\hat{\beta_{k}}(x,y,z) P_{\*M_i}(dx dy dz)$ for $k=1,2,3$. These are compared with similar biomarkers coming from TAC ($bmT_{ki}=TAC_i\hat{\beta}_k $), quantile functions of MIMS from three different axes ($bmadd_{ki}=\int_{0}^{1}Q_{iX}(p)\hat{\beta}_X(p)dp+\int_{0}^{1}Q_{iY}(p)\hat{\beta}_Y(p)dp+\int_{0}^{1}Q_{iZ}(p)\hat{\beta}_Z(p)dp$) and quantile function of composite triaxial MIMS ($bmVM_{ki}=\int_{0}^{1}Q_{iA}(p)\hat{\beta}_A(p)dp$) for $k=1,2,3$. Figure \ref{fig:figbm} displays the scatterplot matrices for the four biomarkers, which are found to be mostly positively correlated for the different cognitive scores. A large amount of spread can be observed in the plots which indicates that these likely capture somewhat different aspects of the association between PA and cognitive functioning.

\begin{figure}[H]
\centering
\includegraphics[width=1\linewidth , height=0.4\linewidth]{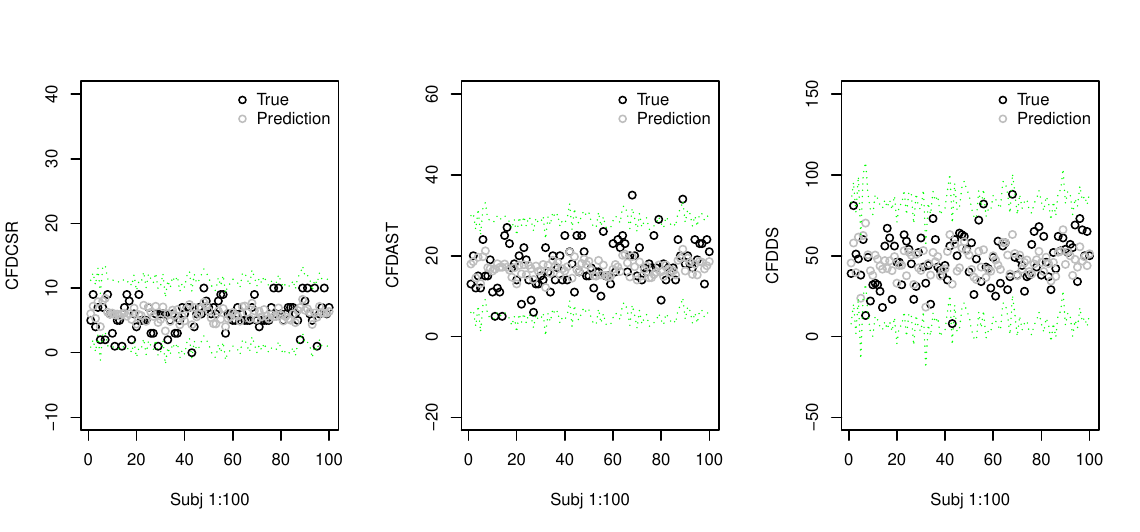}
\caption{Displayed are the prediction intervals (dotted) for the three cognitive scores obtained by the proposed conformal prediction algorithm for test subjects $1,2,\ldots,100$, along with the true and predicted cognitive scores.}
\label{fig:figconf}
\end{figure}

We applied the proposed conformal prediction algorithm to obtain a prediction region with a confidence level of $\alpha=0.05$ for the three cognitive scores of a randomly selected sample of 100 subjects. The estimated coverage from our method is $0.96$, which is close to the nominal coverage. The prediction region is visualized in Figure \ref{fig:figconf} as three prediction intervals corresponding to the three cognitive scores. It can be noticed that a majority of the observed cognitive scores lie within the prediction intervals, as expected. In general, when considering the original scale of the scores, the uncertainty is high. This suggests that additional predictors may be needed to improve prediction accuracy, such as advanced biomarkers related to individual aging or longitudinal history of the subjects. Despite the fact that the new distributional modeling can increase prediction accuracy by more than $20\%$ for different scores, the level of uncertainty remains high. Therefore, caution must be exercised when using the model for personalized interventions. More frequent routine medical tests for longitudinal cognitive capacity characterization may be necessary to create a translational model in practice \citep{hammouri2023uncertainty}.

In conclusion, our proposed MSOMDR framework provides a unified approach for modeling multivariate cognitive outcomes and offers valuable insights into their association with triaxial physical activity data. The framework also enhances interpretability and provides clinically relevant conclusions by quantifying uncertainty in this scientific problem.

\section{Discussion}

\noindent The primary contribution of this paper is the introduction of a novel and general regression framework for analyzing multivariate scalar outcomes based on multidimensional distributional representations as predictors. This framework fills a significant gap in the literature by addressing the statistical challenges associated with working with distributional predictors of higher ($d>1$) dimensions. It overcomes the limitations of traditional summary based or uni-dimensional and univariate scalar-on-distribution regression approaches. For this modelling purpose, we have developed a spline based regularized estimation
approach for modelling the effect of the latent joint density on multivariate response of interest. The new methods is semiparametric, minimizes the impact of curse of dimensionality, and provide clear interpretations in terms of multi-dimensional distributional coefficients on multivariate response of interest.

The use of distributional representation in wearable data analysis is becoming increasingly common, particularly in the field of accelerometer devices \citep{ghosal2022scalar, matabuena2023distributional, doi:10.1177/09622802231192949}. Despite the enhancements in capacity prediction for various applications, the incorporation of multimodal data into these evolving models shows great promise, as we demonstrate in this work. In the context of distributional data analysis \citep{brito2022analysis} and functional compositional analysis \citep{van2014bayes}, our framework, despite its multidistributional nature, offers a key and distinctive feature: the ability to interpret densities as natural predictors. This stands in contrast to prior methods in the literature, which often rely on using the quantile function as predictors or transformations of densities within an unconstrained $L^{2}([0,1])$ space.

Wearable data analysis is of paramount importance because it is essential to measure the statistical association between intensity unit readings from the device and health outcomes. From a technical perspective, quantiles become challenging to generalize in dimensions greater than one. While there may be potential with the notion of depth bands \citep{de2021depth}, it often depends on the specific geometry chosen. Furthermore, the interpretation of quantiles as predictors is not clear, as it creates an individual profile that depends on the specific range of values for that individual. Kernel methods in reproducing kernel Hilbert spaces \citep{matabuena2022kernel, matabuena2023distributional} can be employed but face similar issues of interpretability. Additionally, the application of compositional techniques poses challenges, especially in the context of: i) Multivariate functional data \citep{genest2022orthogonal, hron2022bivariate}; ii) Different support across densities \citep{van2014bayes, wynne2023bayes, petersen2016functional}; iii) The persisting problem of transforming raw functions into standard Hilbert spaces \citep{matabuena2020glucodensities}.

Another distinctive advantage of our modeling strategy is its incorporation of specific techniques for uncertainty quantification based on conformal prediction \citep{vovk2005algorithmic, lei2018distribution}, providing non-asymptotic guarantees. This approach allows for the construction of multivariate prediction regions and offers a deeper understanding of the conditional distribution of the response based on the covariates. Unlike traditional methods that solely focus on the conditional mean, our approach considers the entire conditional distribution. This is particularly advantageous in biomedical applications where the response exhibits high variability and data heterogeneity. In the specific application of cognitive scores modeling, we not only simultaneously predict all three cognitive scores but also estimate predictive limits of the models through uncertainty analysis and address the correlations and interconnections among these parameters.

Our empirical analysis using simulations showcase the accuracy, robustness and advantages of the proposed even when working with finite samples. Furthermore, we demonstrated the application of the proposed method on tri-axial accelerometer data from the National Health and Nutrition Examination Survey (NHANES) 2011-2014 for modelling the association between cognitive scores from three different 
and distributional representation of physical activity among older
adult population. This real-world application served as a compelling demonstration of the significant advantages offered by our approach compared to traditional approaches based on summary level accelerometer metrics. The proposed MSOMDR method captures the dependence between the joint distribution of PA along three different axes and the three cognitive scores in terms of highly interpretable model coefficients. The estimated distributional effects reaffirm the dose-response relationship \citep{bherer2013review,erickson2019physical,zheng2023dose} between physical activity and cognitive function among older adults and 
highlights the importance of high intensity activities which result in higher Y,Z-axis MIMS accumulation for a better cognitive functioning. In contrast to the commonly used summary metrics, such as total activity count derived from marginal distributional representations, or additive distributional approaches, our proposed method also exhibited superior performance in terms of out-of-sample R-squared values, thus highlighting its enhanced predictive capabilities. 



In this article we have used a multi-task learning approach using group penalized regression for estimation of the distributional coefficients, which implicitly introduce correlation among the multivariate outcomes. The proposed framework could be extended to directly accommodate correlation between the outcomes. In this regard a feasible generalized least squares (GLS) type approach \citep{hansen2007generalized} could be pursued for estimation.
Multivariate bootstrap methods can be explored for making inferences about the high dimensional distributional surfaces \citep{eck2018bootstrapping}. Alternatively, Bayesian frameworks \citep{roy2023nonparametric} could be adopted to handle high dimensional distributional predictors \citep{tang2023direct} and multivariate outcomes, which would also aid in uncertainty quantification of the coefficient estimates. 

There are multiple research directions which remain to be explored based on this current work. Within the current proposed framework, as the data dimension $d$ grows ($d>=5$), a fully nonparametric specification and estimation of $\beta(\*s)$ would become computationally challenging. A single index type model specification \citep{hardle1989investigating,ichimura1993semiparametric}. e.g., $\beta(\*s)=\theta(\bm\alpha^T\*s)$, could be an attractive and parsimonious alternative in this regard.
New dimension reduction techniques such as sparse single-index models will be needed to handle the representation of distributional data in high-dimensional spaces more effectively.  These techniques would aim to produce more concise and parsimonious representations of the distributions, enabling efficient analysis and interpretation. Another practical aspect in terms of validation would be to use distributional representations of data from alternative biosensors, such as continuous glucose monitoring or other wearable devices (heart rate, EEG etc).
Considering joint distributions of such multimodal data and their longitudinal changes could provide important scientific insights into human health and human behaviour and will be an interesting area of future research.

\section*{Supplementary Material}
Appendix A, Supplementary Table S1, and Supplementary Figures S1-S3 are available with this paper as Supplementary Material.



\bibliographystyle{chicago}
\bibliography{refs}

\begin{thebibliography}{}

\bibitem[\protect\citeauthoryear{Anderson and McConnell}{Anderson and McConnell}{2007}]{anderson2007cognitive}
Anderson, L.~A. and S.~R. McConnell (2007).
\newblock Cognitive health: an emerging public health issue.
\newblock {\em Alzheimer's \& dementia: the journal of the Alzheimer's Association\/}.

\bibitem[\protect\citeauthoryear{Angelopoulos, Bates, et~al.}{Angelopoulos et~al.}{2023}]{angelopoulos2023conformal}
Angelopoulos, A.~N., S.~Bates, et~al. (2023).
\newblock Conformal prediction: A gentle introduction.
\newblock {\em Foundations and Trends{\textregistered} in Machine Learning\/}~{\em 16\/}(4), 494--591.

\bibitem[\protect\citeauthoryear{Antonsdottir, Low, Chen, Rabinowitz, Yue, Urbanek, Wu, Zeitzer, Rosenberg, Friedman, et~al.}{Antonsdottir et~al.}{2023}]{antonsdottir202324}
Antonsdottir, I.~M., D.~V. Low, D.~Chen, J.~A. Rabinowitz, Y.~Yue, J.~Urbanek, M.~N. Wu, J.~M. Zeitzer, P.~B. Rosenberg, L.~F. Friedman, et~al. (2023).
\newblock 24 h rest/activity rhythms in older adults with memory impairment: Associations with cognitive performance and depressive symptomatology.
\newblock {\em Advanced Biology\/}, e2300138--e2300138.

\bibitem[\protect\citeauthoryear{Bherer, Erickson, Liu-Ambrose, et~al.}{Bherer et~al.}{2013}]{bherer2013review}
Bherer, L., K.~I. Erickson, T.~Liu-Ambrose, et~al. (2013).
\newblock A review of the effects of physical activity and exercise on cognitive and brain functions in older adults.
\newblock {\em Journal of aging research\/}~{\em 2013}.

\bibitem[\protect\citeauthoryear{Breheny and Huang}{Breheny and Huang}{2015a}]{breheny2015group}
Breheny, P. and J.~Huang (2015a).
\newblock Group descent algorithms for nonconvex penalized linear and logistic regression models with grouped predictors.
\newblock {\em Statistics and computing\/}~{\em 25}, 173--187.

\bibitem[\protect\citeauthoryear{Breheny and Huang}{Breheny and Huang}{2015b}]{bre2015}
Breheny, P. and J.~Huang (2015b).
\newblock Group descent algorithms for nonconvex penalized linear and logistic regression models with grouped predictors.
\newblock {\em Statistics and Computing\/}~{\em 25}, 173--187.

\bibitem[\protect\citeauthoryear{Brito and Dias}{Brito and Dias}{2022}]{brito2022analysis}
Brito, P. and S.~Dias (2022).
\newblock {\em Analysis of distributional data}.
\newblock CRC Press.

\bibitem[\protect\citeauthoryear{De~Micheaux, Mozharovskyi, and Vimond}{De~Micheaux et~al.}{2021}]{de2021depth}
De~Micheaux, P.~L., P.~Mozharovskyi, and M.~Vimond (2021).
\newblock Depth for curve data and applications.
\newblock {\em Journal of the American Statistical Association\/}~{\em 116\/}(536), 1881--1897.

\bibitem[\protect\citeauthoryear{Diquigiovanni, Fontana, and Vantini}{Diquigiovanni et~al.}{2022}]{diquigiovanni2022conformal}
Diquigiovanni, J., M.~Fontana, and S.~Vantini (2022).
\newblock Conformal prediction bands for multivariate functional data.
\newblock {\em Journal of Multivariate Analysis\/}~{\em 189}, 104879.

\bibitem[\protect\citeauthoryear{Eck}{Eck}{2018}]{eck2018bootstrapping}
Eck, D.~J. (2018).
\newblock Bootstrapping for multivariate linear regression models.
\newblock {\em Statistics \& Probability Letters\/}~{\em 134}, 141--149.

\bibitem[\protect\citeauthoryear{Erickson, Hillman, Stillman, Ballard, Bloodgood, Conroy, Macko, Marquez, Petruzzello, and Powell}{Erickson et~al.}{2019}]{erickson2019physical}
Erickson, K.~I., C.~Hillman, C.~M. Stillman, R.~M. Ballard, B.~Bloodgood, D.~E. Conroy, R.~Macko, D.~X. Marquez, S.~J. Petruzzello, and K.~E. Powell (2019).
\newblock Physical activity, cognition, and brain outcomes: a review of the 2018 physical activity guidelines.
\newblock {\em Medicine and science in sports and exercise\/}~{\em 51\/}(6), 1242.

\bibitem[\protect\citeauthoryear{Fan, James, and Radchenko}{Fan et~al.}{2015}]{fan2015functional}
Fan, Y., G.~M. James, and P.~Radchenko (2015).
\newblock Functional additive regression.
\newblock {\em The Annals of Statistics\/}~{\em 43\/}(5), 2296--2325.

\bibitem[\protect\citeauthoryear{Genest, Hron, and Ne{\v{s}}lehov{\'a}}{Genest et~al.}{2022}]{genest2022orthogonal}
Genest, C., K.~Hron, and J.~G. Ne{\v{s}}lehov{\'a} (2022).
\newblock Orthogonal decomposition of multivariate densities in bayes spaces and its connection with copulas.
\newblock {\em arXiv preprint arXiv:2206.13898\/}.

\bibitem[\protect\citeauthoryear{Ghosal, Varma, Volfson, Hillel, Urbanek, Hausdorff, Watts, and Zipunnikov}{Ghosal et~al.}{2021}]{gait2020rv}
Ghosal, R., V.~R. Varma, D.~Volfson, I.~Hillel, J.~Urbanek, J.~M. Hausdorff, A.~Watts, and V.~Zipunnikov (2021).
\newblock Distributional data analysis via quantile functions and its application to modelling digital biomarkers of gait in alzheimer’s disease.
\newblock {\em Biostatistics\/}.

\bibitem[\protect\citeauthoryear{Ghosal, Varma, Volfson, Urbanek, Hausdorff, Watts, and Zipunnikov}{Ghosal et~al.}{2022}]{ghosal2022scalar}
Ghosal, R., V.~R. Varma, D.~Volfson, J.~Urbanek, J.~M. Hausdorff, A.~Watts, and V.~Zipunnikov (2022).
\newblock Scalar on time-by-distribution regression and its application for modelling associations between daily-living physical activity and cognitive functions in alzheimer’s disease.
\newblock {\em Scientific reports\/}~{\em 12\/}(1), 11558.

\bibitem[\protect\citeauthoryear{Grundman, Petersen, Ferris, Thomas, Aisen, Bennett, Foster, Jack~Jr, Galasko, Doody, et~al.}{Grundman et~al.}{2004}]{grundman2004mild}
Grundman, M., R.~C. Petersen, S.~H. Ferris, R.~G. Thomas, P.~S. Aisen, D.~A. Bennett, N.~L. Foster, C.~R. Jack~Jr, D.~R. Galasko, R.~Doody, et~al. (2004).
\newblock Mild cognitive impairment can be distinguished from alzheimer disease and normal aging for clinical trials.
\newblock {\em Archives of neurology\/}~{\em 61\/}(1), 59--66.

\bibitem[\protect\citeauthoryear{Hammouri, Mier, F{\'e}lix, Mansournia, Huelin, Casals, and Matabuena}{Hammouri et~al.}{2023}]{hammouri2023uncertainty}
Hammouri, Z. A.~A., P.~R. Mier, P.~F{\'e}lix, M.~A. Mansournia, F.~Huelin, M.~Casals, and M.~Matabuena (2023).
\newblock Uncertainty quantification in medicine science: The next big step.
\newblock {\em Archivos de bronconeumologia\/}, S0300--2896.

\bibitem[\protect\citeauthoryear{Hansen}{Hansen}{2007}]{hansen2007generalized}
Hansen, C.~B. (2007).
\newblock Generalized least squares inference in panel and multilevel models with serial correlation and fixed effects.
\newblock {\em Journal of econometrics\/}~{\em 140\/}(2), 670--694.

\bibitem[\protect\citeauthoryear{H{\"a}rdle and Stoker}{H{\"a}rdle and Stoker}{1989}]{hardle1989investigating}
H{\"a}rdle, W. and T.~M. Stoker (1989).
\newblock Investigating smooth multiple regression by the method of average derivatives.
\newblock {\em Journal of the American statistical Association\/}~{\em 84\/}(408), 986--995.

\bibitem[\protect\citeauthoryear{Henry, Crawford, and Phillips}{Henry et~al.}{2004}]{henry2004verbal}
Henry, J.~D., J.~R. Crawford, and L.~H. Phillips (2004).
\newblock Verbal fluency performance in dementia of the alzheimer’s type: a meta-analysis.
\newblock {\em Neuropsychologia\/}~{\em 42\/}(9), 1212--1222.

\bibitem[\protect\citeauthoryear{Hron, Machalov{\'a}, and Menafoglio}{Hron et~al.}{2022}]{hron2022bivariate}
Hron, K., J.~Machalov{\'a}, and A.~Menafoglio (2022).
\newblock Bivariate densities in bayes spaces: orthogonal decomposition and spline representation.
\newblock {\em Statistical Papers\/}, 1--39.

\bibitem[\protect\citeauthoryear{Hron, Menafoglio, Templ, Hruzova, and Filzmoser}{Hron et~al.}{2016}]{hron2016simplicial}
Hron, K., A.~Menafoglio, M.~Templ, K.~Hruzova, and P.~Filzmoser (2016).
\newblock Simplicial principal component analysis for density functions in bayes spaces.
\newblock {\em Computational Statistics \& Data Analysis\/}~{\em 94}, 330--350.

\bibitem[\protect\citeauthoryear{Ichimura}{Ichimura}{1993}]{ichimura1993semiparametric}
Ichimura, H. (1993).
\newblock Semiparametric least squares (sls) and weighted sls estimation of single-index models.
\newblock {\em Journal of econometrics\/}~{\em 58\/}(1-2), 71--120.

\bibitem[\protect\citeauthoryear{Janssen, Clarke, Carson, Chaput, Giangregorio, Kho, Poitras, Ross, Saunders, Ross-White, et~al.}{Janssen et~al.}{2020}]{janssen2020systematic}
Janssen, I., A.~E. Clarke, V.~Carson, J.-P. Chaput, L.~M. Giangregorio, M.~E. Kho, V.~J. Poitras, R.~Ross, T.~J. Saunders, A.~Ross-White, et~al. (2020).
\newblock A systematic review of compositional data analysis studies examining associations between sleep, sedentary behaviour, and physical activity with health outcomes in adults.
\newblock {\em Applied physiology, nutrition, and metabolism\/}~{\em 45\/}(10), S248--S257.

\bibitem[\protect\citeauthoryear{Javed, Sarwar, Khan, Iwendi, Mittal, and Kumar}{Javed et~al.}{2020}]{javed2020analyzing}
Javed, A.~R., M.~U. Sarwar, S.~Khan, C.~Iwendi, M.~Mittal, and N.~Kumar (2020).
\newblock Analyzing the effectiveness and contribution of each axis of tri-axial accelerometer sensor for accurate activity recognition.
\newblock {\em Sensors\/}~{\em 20\/}(8), 2216.

\bibitem[\protect\citeauthoryear{Jašková, Palarea-Albaladejo, Gába, Dumuid, Željko Pedišić, Pelclová, and Hron}{Jašková et~al.}{0}]{doi:10.1177/09622802231192949}
Jašková, P., J.~Palarea-Albaladejo, A.~Gába, D.~Dumuid, Željko Pedišić, J.~Pelclová, and K.~Hron (0).
\newblock Compositional functional regression and isotemporal substitution analysis: Methods and application in time-use epidemiology.
\newblock {\em Statistical Methods in Medical Research\/}~{\em 0\/}(0), 09622802231192949.
\newblock PMID: 37590096.

\bibitem[\protect\citeauthoryear{John, Tang, Albinali, and Intille}{John et~al.}{2019}]{john2019open}
John, D., Q.~Tang, F.~Albinali, and S.~Intille (2019).
\newblock An open-source monitor-independent movement summary for accelerometer data processing.
\newblock {\em Journal for the Measurement of Physical Behaviour\/}~{\em 2\/}(4), 268--281.

\bibitem[\protect\citeauthoryear{Lei, G’Sell, Rinaldo, Tibshirani, and Wasserman}{Lei et~al.}{2018}]{lei2018distribution}
Lei, J., M.~G’Sell, A.~Rinaldo, R.~J. Tibshirani, and L.~Wasserman (2018).
\newblock Distribution-free predictive inference for regression.
\newblock {\em Journal of the American Statistical Association\/}~{\em 113\/}(523), 1094--1111.

\bibitem[\protect\citeauthoryear{Li and Liu}{Li and Liu}{2008}]{li2008multivariate}
Li, J. and R.~Y. Liu (2008).
\newblock Multivariate spacings based on data depth: I. construction of nonparametric multivariate tolerance regions.

\bibitem[\protect\citeauthoryear{Marx and Eilers}{Marx and Eilers}{2005}]{marx2005multidimensional}
Marx, B.~D. and P.~H. Eilers (2005).
\newblock Multidimensional penalized signal regression.
\newblock {\em Technometrics\/}~{\em 47\/}(1), 13--22.

\bibitem[\protect\citeauthoryear{Matabuena, Felix, Garcia-Meixide, and Gude}{Matabuena et~al.}{2022}]{matabuena2022kernel}
Matabuena, M., P.~Felix, C.~Garcia-Meixide, and F.~Gude (2022).
\newblock Kernel machine learning methods to handle missing responses with complex predictors. application in modelling five-year glucose changes using distributional representations.
\newblock {\em Computer Methods and Programs in Biomedicine\/}~{\em 221}, 106905.

\bibitem[\protect\citeauthoryear{Matabuena and Petersen}{Matabuena and Petersen}{2021}]{matabuena2021distributional}
Matabuena, M. and A.~Petersen (2021).
\newblock Distributional data analysis with accelerometer data in a nhanes database with nonparametric survey regression models.
\newblock {\em arXiv\/}.

\bibitem[\protect\citeauthoryear{Matabuena and Petersen}{Matabuena and Petersen}{2023}]{matabuena2023distributional}
Matabuena, M. and A.~Petersen (2023).
\newblock Distributional data analysis of accelerometer data from the nhanes database using nonparametric survey regression models.
\newblock {\em Journal of the Royal Statistical Society Series C: Applied Statistics\/}~{\em 72\/}(2), 294--313.

\bibitem[\protect\citeauthoryear{Matabuena, Petersen, Vidal, and Gude}{Matabuena et~al.}{2021}]{matabuena2020glucodensities}
Matabuena, M., A.~Petersen, J.~C. Vidal, and F.~Gude (2021).
\newblock Glucodensities: a new representation of glucose profiles using distributional data analysis.
\newblock {\em Statistical Methods in Medical Research\/}~{\em 30\/}(6), 1445--1464.

\bibitem[\protect\citeauthoryear{Meunier, Pontil, and Ciliberto}{Meunier et~al.}{2022}]{meunier2022distribution}
Meunier, D., M.~Pontil, and C.~Ciliberto (2022).
\newblock Distribution regression with sliced wasserstein kernels.
\newblock In {\em International Conference on Machine Learning}, pp.\  15501--15523. PMLR.

\bibitem[\protect\citeauthoryear{Papadopoulos, Proedrou, Vovk, and Gammerman}{Papadopoulos et~al.}{2002}]{papadopoulos2002inductive}
Papadopoulos, H., K.~Proedrou, V.~Vovk, and A.~Gammerman (2002).
\newblock Inductive confidence machines for regression.
\newblock In {\em Machine Learning: ECML 2002: 13th European Conference on Machine Learning Helsinki, Finland, August 19--23, 2002 Proceedings 13}, pp.\  345--356. Springer.

\bibitem[\protect\citeauthoryear{Pecanka, Van Der~Vaart, and Jonker}{Pecanka et~al.}{2019}]{pecanka2019modeling}
Pecanka, J., A.~Van Der~Vaart, and M.~Jonker (2019).
\newblock Modeling association between multivariate correlated outcomes and high-dimensional sparse covariates: the adaptive svs method.
\newblock {\em Journal of Applied Statistics\/}~{\em 46\/}(5), 893--913.

\bibitem[\protect\citeauthoryear{Petersen and M{\"u}ller}{Petersen and M{\"u}ller}{2016}]{petersen2016functional}
Petersen, A. and H.-G. M{\"u}ller (2016).
\newblock Functional data analysis for density functions by transformation to a hilbert space.

\bibitem[\protect\citeauthoryear{Petersen and M{\"u}ller}{Petersen and M{\"u}ller}{2019}]{petersen2019frechet}
Petersen, A. and H.-G. M{\"u}ller (2019).
\newblock Fr{\'e}chet regression for random objects with euclidean predictors.

\bibitem[\protect\citeauthoryear{Petersen, Zhang, and Kokoszka}{Petersen et~al.}{2021}]{petersen2021modeling}
Petersen, A., C.~Zhang, and P.~Kokoszka (2021).
\newblock Modeling probability density functions as data objects.
\newblock {\em Econometrics and Statistics\/}.

\bibitem[\protect\citeauthoryear{Proust-Lima, Amieva, Dartigues, and Jacqmin-Gadda}{Proust-Lima et~al.}{2007}]{proust2007sensitivity}
Proust-Lima, C., H.~Amieva, J.-F. Dartigues, and H.~Jacqmin-Gadda (2007).
\newblock Sensitivity of four psychometric tests to measure cognitive changes in brain aging-population--based studies.
\newblock {\em American journal of epidemiology\/}~{\em 165\/}(3), 344--350.

\bibitem[\protect\citeauthoryear{{R Core Team}}{{R Core Team}}{2018}]{Rsoft}
{R Core Team} (2018).
\newblock {\em R: A Language and Environment for Statistical Computing}.
\newblock Vienna, Austria: R Foundation for Statistical Computing.

\bibitem[\protect\citeauthoryear{Ramsay and Silverman}{Ramsay and Silverman}{2005}]{Ramsay05functionaldata}
Ramsay, J. and B.~Silverman (2005).
\newblock {\em Functional Data Analysis}.
\newblock New York: Springer-Verlag.

\bibitem[\protect\citeauthoryear{Reiss, Goldsmith, Shang, and Ogden}{Reiss et~al.}{2017}]{reiss2017methods}
Reiss, P.~T., J.~Goldsmith, H.~L. Shang, and R.~T. Ogden (2017).
\newblock Methods for scalar-on-function regression.
\newblock {\em International Statistical Review\/}~{\em 85\/}(2), 228--249.

\bibitem[\protect\citeauthoryear{Romano, Patterson, and Candes}{Romano et~al.}{2019}]{romano2019conformalized}
Romano, Y., E.~Patterson, and E.~Candes (2019).
\newblock Conformalized quantile regression.
\newblock {\em Advances in neural information processing systems\/}~{\em 32}.

\bibitem[\protect\citeauthoryear{Roy}{Roy}{2023}]{roy2023nonparametric}
Roy, A. (2023).
\newblock Nonparametric group variable selection with multivariate response for connectome-based modelling of cognitive scores.
\newblock {\em Journal of the Royal Statistical Society Series C: Applied Statistics\/}~{\em 72\/}(4), 872--888.

\bibitem[\protect\citeauthoryear{Sherr, Cengiz, Palerm, Clark, Kurtz, Roy, Carria, Cantwell, Tamborlane, and Weinzimer}{Sherr et~al.}{2013}]{sherr2013reduced}
Sherr, J.~L., E.~Cengiz, C.~C. Palerm, B.~Clark, N.~Kurtz, A.~Roy, L.~Carria, M.~Cantwell, W.~V. Tamborlane, and S.~A. Weinzimer (2013).
\newblock Reduced hypoglycemia and increased time in target using closed-loop insulin delivery during nights with or without antecedent afternoon exercise in type 1 diabetes.
\newblock {\em Diabetes care\/}~{\em 36\/}(10), 2909--2914.

\bibitem[\protect\citeauthoryear{Talsk{\'a}, Hron, and Grygar}{Talsk{\'a} et~al.}{2021}]{talska2021compositional}
Talsk{\'a}, R., K.~Hron, and T.~M. Grygar (2021).
\newblock Compositional scalar-on-function regression with application to sediment particle size distributions.
\newblock {\em Mathematical Geosciences\/}, 1--29.

\bibitem[\protect\citeauthoryear{Tang, Zhao, Caffo, and Datta}{Tang et~al.}{2023}]{tang2023direct}
Tang, B., Y.~Zhao, B.~Caffo, and A.~Datta (2023).
\newblock Direct bayesian regression for distribution-valued covariates.
\newblock {\em arXiv preprint arXiv:2303.06434\/}.

\bibitem[\protect\citeauthoryear{Tang, Zhao, Venkataraman, Tsapkini, Lindquist, Pekar, and Caffo}{Tang et~al.}{2023}]{tang2023differences}
Tang, B., Y.~Zhao, A.~Venkataraman, K.~Tsapkini, M.~A. Lindquist, J.~Pekar, and B.~Caffo (2023).
\newblock Differences in functional connectivity distribution after transcranial direct-current stimulation: A connectivity density point of view.
\newblock {\em Human Brain Mapping\/}~{\em 44\/}(1), 170--185.

\bibitem[\protect\citeauthoryear{Van~den Boogaart, Egozcue, and Pawlowsky-Glahn}{Van~den Boogaart et~al.}{2014}]{van2014bayes}
Van~den Boogaart, K.~G., J.~J. Egozcue, and V.~Pawlowsky-Glahn (2014).
\newblock Bayes hilbert spaces.
\newblock {\em Australian \& New Zealand Journal of Statistics\/}~{\em 56\/}(2), 171--194.

\bibitem[\protect\citeauthoryear{Varma, Ghosal, Hillel, Volfson, Weiss, Urbanek, Hausdorff, Zipunnikov, and Watts}{Varma et~al.}{2021}]{gait2020vr}
Varma, V.~R., R.~Ghosal, I.~Hillel, D.~Volfson, J.~Weiss, J.~Urbanek, J.~M. Hausdorff, V.~Zipunnikov, and A.~Watts (2021).
\newblock Continuous gait monitoring discriminates community dwelling mild ad from cognitively normal controls.
\newblock {\em Alzheimer's \& Dementia: Translational Research \& Clinical Interventions\/}~{\em 7\/}(1), e12131.

\bibitem[\protect\citeauthoryear{Vovk, Gammerman, and Shafer}{Vovk et~al.}{2005}]{vovk2005algorithmic}
Vovk, V., A.~Gammerman, and G.~Shafer (2005).
\newblock {\em Algorithmic learning in a random world}, Volume~29.
\newblock Springer.

\bibitem[\protect\citeauthoryear{Wynne}{Wynne}{2023}]{wynne2023bayes}
Wynne, G. (2023).
\newblock Bayes hilbert spaces for posterior approximation.
\newblock {\em arXiv preprint arXiv:2304.09053\/}.

\bibitem[\protect\citeauthoryear{Young and Mathew}{Young and Mathew}{2020}]{young2020nonparametric}
Young, D.~S. and T.~Mathew (2020).
\newblock Nonparametric hyperrectangular tolerance and prediction regions for setting multivariate reference regions in laboratory medicine.
\newblock {\em Statistical Methods in Medical Research\/}~{\em 29\/}(12), 3569--3585.

\bibitem[\protect\citeauthoryear{Zhang}{Zhang}{2010}]{zhang2010nearly}
Zhang, C.-H. (2010).
\newblock Nearly unbiased variable selection under minimax concave penalty.
\newblock {\em The Annals of Statistics\/}~{\em 38\/}(2), 894--942.

\bibitem[\protect\citeauthoryear{Zheng, Pleuss, Turner, Ducharme, and Aguiar}{Zheng et~al.}{2023}]{zheng2023dose}
Zheng, P., J.~D. Pleuss, D.~S. Turner, S.~W. Ducharme, and E.~J. Aguiar (2023).
\newblock Dose--response association between physical activity (daily mims, peak 30-minute mims) and cognitive function among older adults: Nhanes 2011--2014.
\newblock {\em The Journals of Gerontology: Series A\/}~{\em 78\/}(2), 286--291.

\bibitem[\protect\citeauthoryear{Zhu and Cao}{Zhu and Cao}{2021}]{zhu2021distributional}
Zhu, J. and J.~Cao (2021).
\newblock Distributional representation of resting-state fmri for functional brain connectivity analysis.
\newblock {\em Neurocomputing\/}~{\em 427}, 156--168.

\end{thebibliography}


\begin{thebibliography}{}

\bibitem[\protect\citeauthoryear{Kuchibhotla}{Kuchibhotla}{2020}]{kuchibhotla2020exchangeability}
Kuchibhotla, A.~K. (2020).
\newblock Exchangeability, conformal prediction, and rank tests.
\newblock {\em arXiv preprint arXiv:2005.06095\/}.

\end{thebibliography}
\end{document}


\def\spacingset#1{\renewcommand{\baselinestretch}%
{#1}\small\normalsize} \spacingset{1}


\if0\blind
{
  \title{\bf Supplementary Material for Multivariate Scalar on Multidimensional Distribution Regression}
 \author{Rahul Ghosal$^{1,\ast}$, Marcos Matabuena$^{2,3}$ \\
  \\
$^{1}$ Department of Epidemiology and Biostatistics, University of South Carolina \\
$^{2}$ Department of Biostatistics, Harvard University\\
$^{3}$ University of Santiago de Compostela, A Coruña, Spain\\
}
  \maketitle
} \fi

\if1\blind
{
  \bigskip
  \bigskip
  \bigskip
  \begin{center}
    {\LARGE\bf Title}
\end{center}
  \medskip
} \fi

\bigskip

\vfill

\newpage
\spacingset{1.5} 

\section{Appendix A: Theory}
We  present the marginal finite sample guarantee of the uncertainty quantification defined in Algorithm 1. To support our analysis, we first introduce several technical results.

\begin{definition}
\noindent Random variables $W_{1},\dots, W_{n}$ for $n\geq 1$  are said to be exchangeable if

\begin{equation}
(W_{1},\dots,W_{n}) \overset{d}{=} (W_{\pi(1)},\dots,W_{\pi(n)})
\end{equation}
\noindent for any permutation $\pi: [n]\to [n]$.  Intuitively, exchangeability means that the index of the random variables is immaterial.
\end{definition}

\begin{definition} For a set of the real numbers $\mathcal{S}= \{x_1,\dots,x_n\}$, define the rank of $x_i$ among $\mathcal{S}$ as
	
	\begin{equation}
rank(x_i; \mathcal{S})= |j\in [n]: x_{j}+\epsilon U_{j}\leq x_{i}+ \epsilon U_{i}|,
	\end{equation}
	
where $\epsilon>0$ is arbitrary $U_{1},\dots, U_{n}$ are iid $U[-1,1]$ random variable.	
\end{definition}

\begin{theorem}\label{theorem:int}
\cite{kuchibhotla2020exchangeability}, if $W_{1},\dots W_{n}$ are exchangeable random variables, then for any $\epsilon>0$,
	
	\begin{equation}
rank(x_i; \{W_{1},\dots W_{n}\}: i\in [n]) \sim  Unif(\{\pi: [n]\to [n]\}).
	\end{equation}
	
	\noindent Here $Unif(\{\pi: [n]\to [n]\})$ represents the uniform distribution over all permutation of $[n]$, that is, each permutation has an equal probability of $\frac{1}{n!}$.
\end{theorem}
	
	\begin{corollary} \label{corol:uniform}  Under the assumptions of Theorem \ref{theorem:int}, for any $\epsilon>0$, we have
\begin{equation}
\mathbb{P}(rank(x_i; \{W_{1},\dots W_{n}\})\leq t)= \frac{[t]}{n}
\end{equation}

\noindent where, for $t\in \mathbb{R}$, $[t]$ represents  the largest integer value than or equal to $t$. Moreover, the random variable 	$P:= rank(x_i; \{W_{1},\dots W_{n}\})\leq t)/n$ is a valid $p-value$, i.e,

\begin{equation}
\mathbb{P}(P\leq \alpha)\leq \alpha \text{ for all }   \alpha\in [0,1].
\end{equation}
\end{corollary}

\begin{proof}(Proposition $1$)
We begin by noting in our algorithm definition that we split $\mathcal{D}_{n} = \mathcal{D}_{\text{train}1} \cup \mathcal{D}_{\text{train}2} \cup \mathcal{D}_{\text{calibration}}$ into three disjoint sets, and the random elements of $\mathcal{D}_{n}$ are independent and identically distributed with respect to $(\mathbf{X}, P_{\mathbf{Z}},\*Y)$. Therefore, the random elements of $\mathcal{D}_{n}$ are exchangeable.

The estimators $\hat{m}(\cdot)$ and $\hat{s}(\cdot)$ are computed using $\mathcal{D}_{\text{train}1}$ and $\mathcal{D}_{\text{train}2}$, respectively. Then, the random elements from $\mathcal{D}_{\text{calibration}}$ are exchangeable if we conditioned on $\mathcal{D}_{\text{train}1} \cup \mathcal{D}_{\text{train}2}$, since $\hat{m}(\cdot)$ and $\hat{s}(\cdot)$ are fixed functions. Consequently, the sequence $\{R_{i}\}_{i\in \mathcal{D}_{\text{calibration}}}$ is exchangeable.

Now, as we are estimating the empirical quantile $1-\alpha$ from $\{R_{i}\}_{i\in \mathcal{D}_{\text{calibration}}}$ (essentially a surrogate for a rank representation) and for the exchaengability property,  by construction of prediction region and interpretation of empirical quantile
$\hat{q}_{1-\alpha}$ (as the radius of the interval from standardized residuals across the dimensions),  we can apply Corollary $1$ to obtain the desired result about the marginal coverage of the proposed prediction region, that is,

     \begin{equation*}
P(\*Y\in \widehat{\mathcal{C}}_{n}^{\alpha}(\*X,P_{\*Z}))\geq 1-\alpha.
\end{equation*}
    
\end{proof}

\section{Supplementary Table}
\begin{table}[H]
\centering
\caption{Descriptive statistics for the complete, male and female samples in the NHANES application. The p-values are from two-sample t-test.}
\vspace{3 mm}
\label{tab:my-tabler}
\centering
\small
\begin{tabular}{cccccccc}
\hline
Characteristic     & \multicolumn{1}{c|}{Complete (n=1947)} & \multicolumn{1}{c|}{Male (n=893)} & \multicolumn{1}{c|}{Female (n=1054)} & p-value         \\ \hline
                   & Mean(sd)          & Mean(sd)     & Mean(sd)            &                 \\ \hline
Age                & 70.1 (6.8)                        &70.2 (6.8)            & 70.0 (60.7)               & 0.50            \\ \hline
CFDCSR  & 5.9 (2.3)                       & 5.5 (2.2)           & 6.3 (2.3)                & $<0.0001$          \\ 
\hline
CFDAST & 16.6 (5.5)                       & 16.9 (5.5)           & 16.5 (5.4)               &   0.10       \\ 
\hline
CFDDS  & 45.7 (16.8)                       & 43.2 (15.6)           & 47.8 (17.5)                &   $<0.0001$       \\ 
\hline
\end{tabular}

\end{table}

\section{Supplementary Figure}

\begin{figure}[H]
\centering
\includegraphics[width=1\linewidth , height=0.9\linewidth]{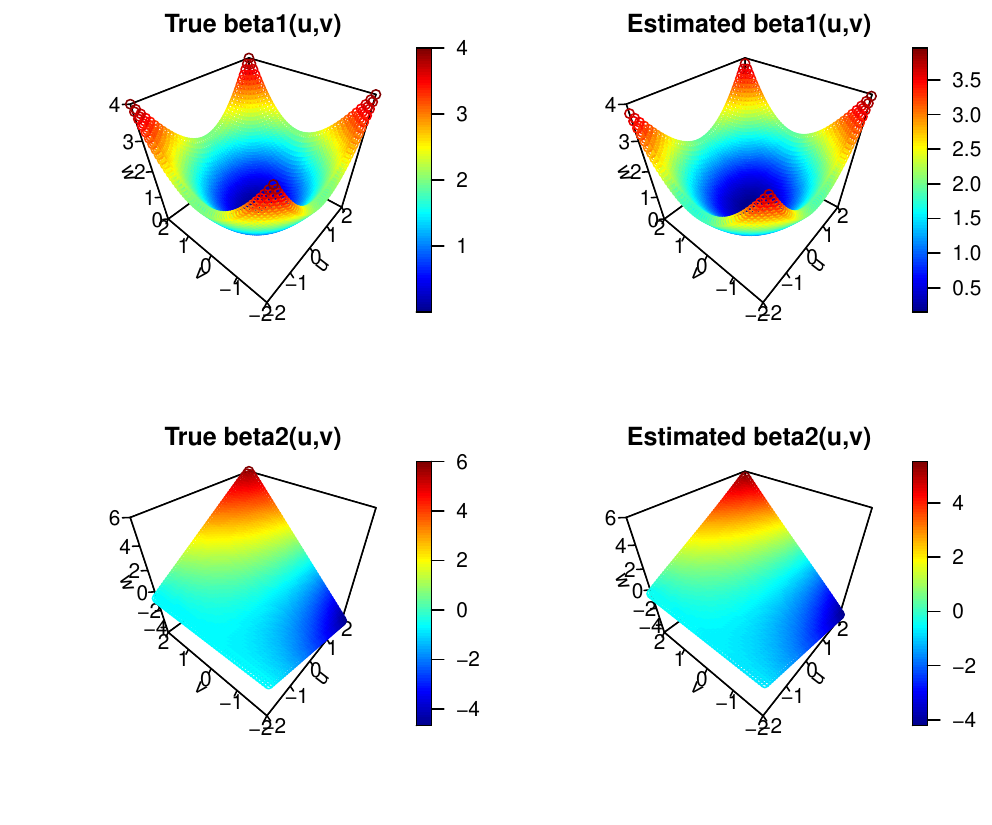}
\caption{Displayed are the true (left) and Monte Carlo mean of the estimated distributional effects (right) $\beta_1(u,v)$,$\beta_2(u,v)$, scenario A1, n=500.}
\label{fig:fig21}
\end{figure}

\begin{figure}[H]
\centering
\includegraphics[width=1\linewidth , height=0.9\linewidth]{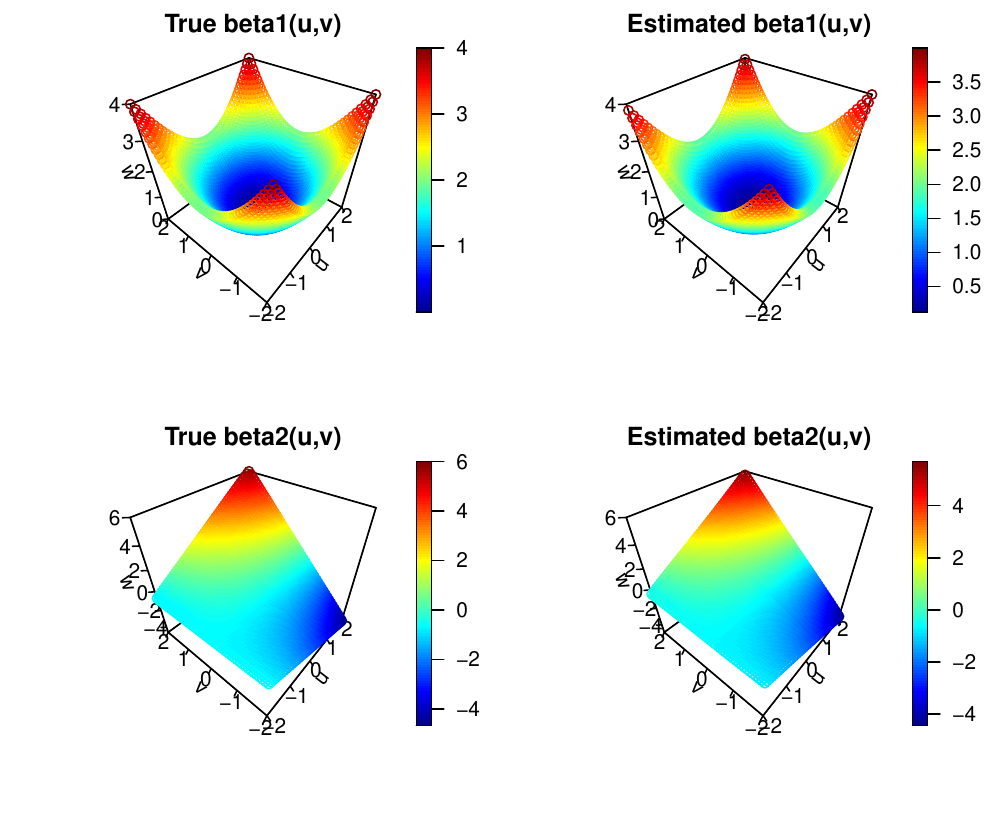}
\caption{Displayed are the true (left) and Monte Carlo mean of the estimated distributional effects (right) $\beta_1(u,v)$,$\beta_2(u,v)$, scenario A1, n=2000.}
\label{fig:fig22}
\end{figure}

\begin{figure}[H]
\centering
\includegraphics[width=1\linewidth , height=0.65\linewidth]{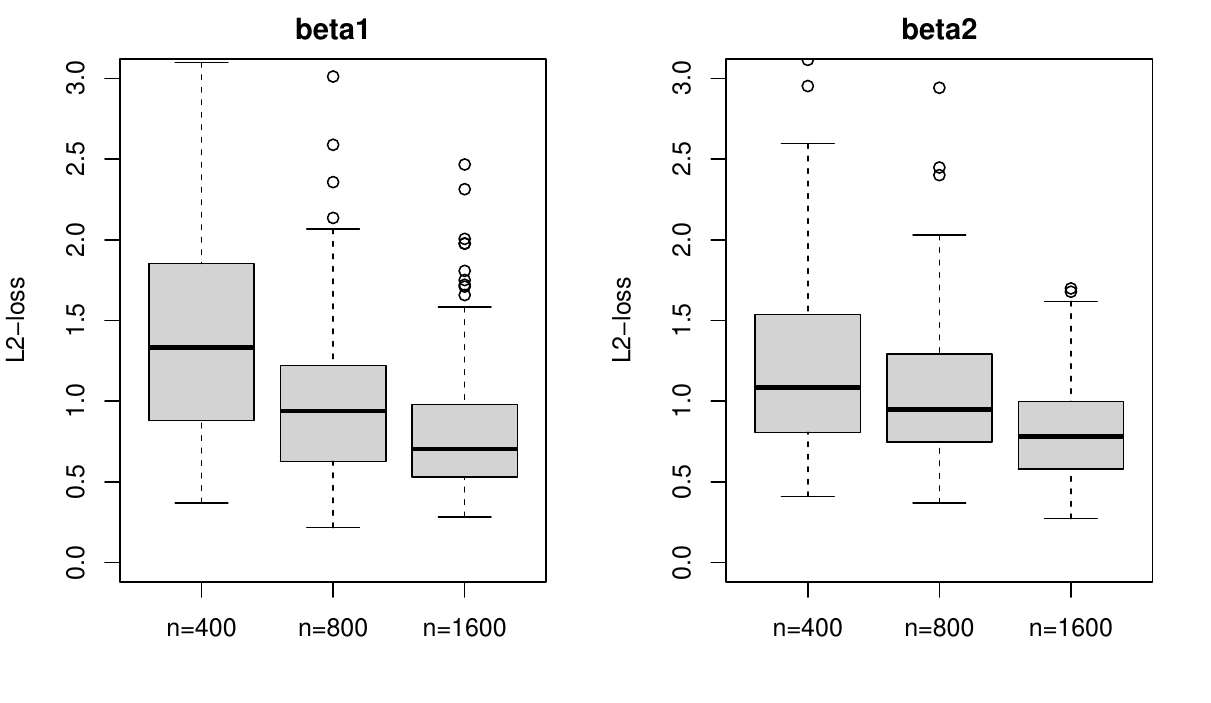}
\caption{Displayed are the distribution of $L^2$ loss between the true and estimated distributional coefficients $\beta_1(u,v)$,$\beta_2(u,v)$, across 100 MC replications, Scenario A1.}
\label{fig:fig10}
\end{figure}

\bibliographystyle{chicago}
\bibliography{refs}